\documentclass{article}
\usepackage{bookstyle}
\usepackage{graphicx}
\usepackage{cite}
\begin{document}
\author{V.~E.~Kravtsov}
\address{The Abdus Salam International Centre for Theoretical Physics,
Strada Costiera 11, 34100 Trieste, Italy.
\\Landau Institute for Theoretical 
Physics,
2 Kosygina Street, 117940 Moscow, 
Russia}
\title{Non-linear Quantum Coherence Effects in Driven Mesoscopic Systems}
\photo{\includegraphics[width=80mm]{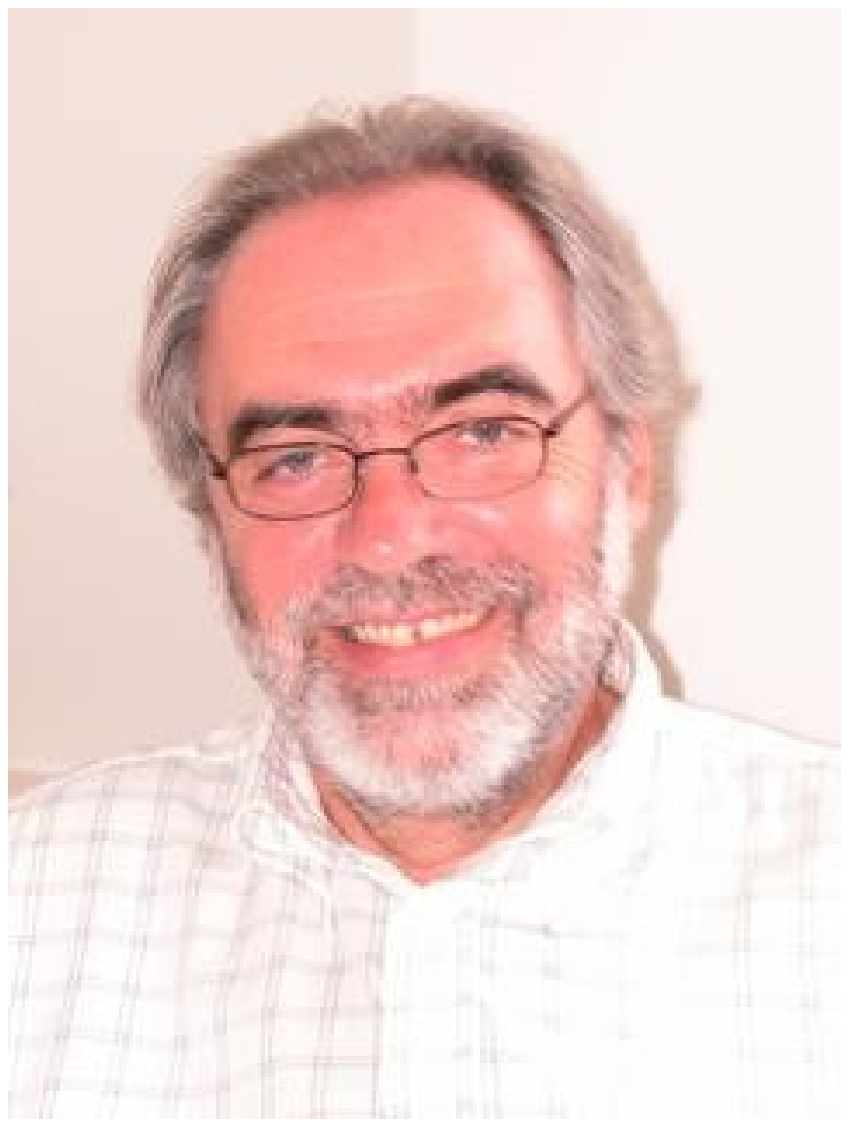}
}
\frontmatter
\maketitle    
\mainmatter

\section{Introduction}
Over the last two decades  theory and experiment on quantum
disordered and chaotic systems have been an extremely successful
field of research. The milestones on this rout were:
\begin{itemize}

\item
Weak Anderson localization in disordered metals
\item
Universal conductance fluctuations
\item
Application of random matrix theory to quantum disordered and
chaotic systems
\end{itemize}
However the mainstream of research was so far limited to a {\it
linear} response of quantum systems where the kinetic and thermodynamic
properties are calculated essentially at equilibrium. 
The main goal of this course
of lectures is to develop a theory of {\it nonlinear} response to
a time-dependent perturbation in  the same way as the
theory of Anderson localization and mesoscopic phenomena. 
It will require an extension of an existing methods to 
{\it non-equilibrium
phenomena}.

The four lectures will include the following topics:
\begin{itemize}
\item
Perturbative theory of weak Anderson localization
\item
Keldysh formulation of nonlinear response theory
\item
Mesoscopic rings under AC pumping and  quantum rectification
\item
Theory of weak dynamic localization in quantum dots
\end{itemize}
\section{Weak Anderson localization in disordered systems}
In this section we consider the main steps in the calculus of weak
localization theory \cite{AGD, GLH}. The main object to study is the
frequency-dependent conductivity
\begin{eqnarray}
\label{cond} \sigma_{\alpha
\beta}(\omega)&=&\int_{-\infty}^{+\infty}\frac{d\varepsilon}{4\pi\omega}
\left[f(\varepsilon)-f(\varepsilon-\omega) \right]\\ \nonumber
&\times& \int\frac{d{\bf r}d{\bf r'}}{Vol}\,\langle
e\hat{v}_{\alpha}(G^{R}-G^{A})_{{\bf r},{\bf
r'};\varepsilon}\;e\hat{v}_{\beta}(G^{R}-G^{A})_{{\bf r'},{\bf
r};\varepsilon-\omega} \rangle
\end{eqnarray}
where $G_{{\bf r'},{\bf r};\varepsilon}^{R/A}$ are
retarded (advanced) electron Green's functions, $f(\varepsilon)$ is
the Fermi energy distribution function and $\hat{v}_{\alpha}$ is
the velocity operator . One can convince oneself using the
representation in terms of exact eigenfunctions $\Psi_{n}({\bf
r})$ and exact eigenvalues $E_{n}$
\begin{eqnarray}
\label{GF} G_{{\bf r'},{\bf
r};\varepsilon}^{R/A}=\sum_{n}\frac{\Psi_{n}({\bf
r})\Psi^{*}_{n}({\bf r' })}{\varepsilon-E_{n}\pm i0}
\end{eqnarray}
that Eq.(\ref{cond}) reduces to a familiar Fermi Golden rule
expression:
\begin{equation}
\label{FGR}
\sigma_{\alpha\beta}(\omega)=2\pi 
\sum_{E_{n}>E_{m}} \langle n|J_{\alpha}|m\rangle \langle 
m|J_{\beta}|n\rangle\,\delta(E_{n}-E_{m}-\omega)\;\frac{f_{E_{m}}-f_{E_{n}}}{
E_{n}-E_{m}} 
\end{equation}
Eq.(\ref{GF}) is convenient to prove exact identities
but is useless for practical calculus. In disordered systems with
the momentum relaxation time $\tau$ the following expression for
the disorder average Green's function is very useful:
\begin{equation}
\label{GFu} \langle
G^{R/A}\rangle_{\varepsilon}=\frac{1}{\varepsilon-\xi_{{\bf
p}}\pm\frac{i}{2\tau}}.
\end{equation}
where $\xi_{{\bf p}}=\varepsilon({\bf p})-E_{F}$ is related with
the electron dispersion law $\varepsilon({\bf p})$ relative the
Fremi energy $E_{F}$. In what follows we will assume $\xi=\xi_{p}$
depending only on $|{\bf p}|$. Then $\xi$ and the unit vector of
the direction of electron momentum  ${\bf n}$ constitute a
convenient variables of integration over ${\bf p}$:
\begin{equation}
\label{int} \int ... d{\bf p}\rightarrow \int \,\nu(\xi) d\xi\int
d{\bf n} ...
\end{equation}
For a typical metal with $E_{F}$ far from the band edges and the
energy scale of interest $\varepsilon << E_{F}$ the density of
states $\nu(\xi)$ can be approximated by a constant  $\nu=\nu(0)$ at
the Fermi level and the integration over $\xi$ can be extended
over the entire real axis $(-\infty, +\infty)$.
\subsection{Drude approximation}
The simplest diagrams for the frequency-dependent conductivity are
shown in Fig.1.
 \begin{figure}[h]
%\fbox{\vtop to5cm{\vss\hsize=.9\hsize\vglue
%0cm\hspace{0.01\hsize}\hskip -.5cm
%\hspace{0.1\hsize}\vss}}
\includegraphics[width=110mm]{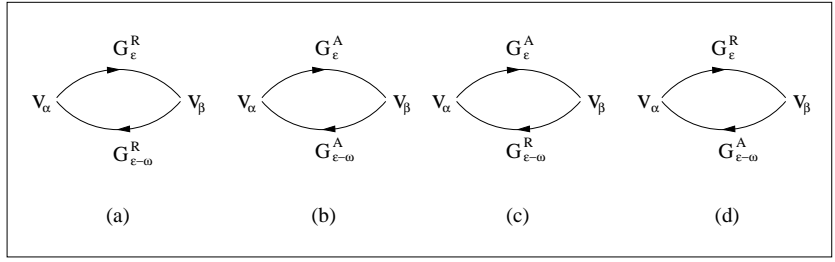}
\caption{ Diagrams for the Drude conductivity.} \label{Drude}
\end{figure}
The diagrams Fig.1a,b are given by the integrals
\begin{equation}
\label{1a} \int d{\bf
n}\,v_{\alpha}v_{\beta}\int_{-\infty}^{+\infty}\frac{d\xi}
{(\varepsilon-\xi\pm\frac{i}{2\tau})(\varepsilon-\omega-\xi\pm\frac{i}{2\tau})}=0
\end{equation}
where $v_{\alpha}\approx v_{F}n_{\alpha}$.

It is important that the integral over $\xi$ is the pole integral
with all poles in the same complex half-plane. This is why such
integrals are equal to zero. In contrast to that the diagrams of
Fig.1c,d that contain both retarded ($G^{R}$) and advanced
($G^{A}$) Green's functions correspond to a similar integral with
poles lying in {\it different} half-planes of $\xi$. Such an
integral over $\xi$ can be done immediately using the residue
theorem:
\begin{equation}
\label{1c} \int_{-\infty}^{+\infty}\frac{d\xi}
{(\varepsilon-\xi\pm\frac{i}{2\tau})(\varepsilon-\omega-\xi\mp\frac{i}{2\tau})}=
\frac{2\pi\tau}{(1-i\omega\tau)}.
\end{equation}
The angular integral is trivial:
\begin{equation}
\label{ang}
 \int d{\bf
n}\,v_{\alpha}v_{\beta}=v_{F}^{2}\,\frac{\delta_{\alpha\beta}}{d},
\end{equation}
where $d$ is the dimensionality of space.

Now all what we need to compute $\sigma_{\alpha\beta}$ using the
simplest diagrams of Fig.1 is the identity:
\begin{equation}
\label{id}
\int_{-\infty}^{+\infty}\left[f(\varepsilon)-f(\varepsilon-\omega)
\right]\,d\varepsilon=-\omega.
\end{equation}
This identity holds for all functions $f(\varepsilon)$ obeying the
Fermi boundary conditions $f(\varepsilon)\rightarrow 0$ at
$\varepsilon\rightarrow +\infty$ and $f(\varepsilon)\rightarrow 1$
at $\varepsilon\rightarrow -\infty$.

The result of the calculations is the Drude conductivity:
\begin{equation}
\label{DR} \sigma^{(D)}_{\alpha\beta}=\frac{e^2 \nu
D_{0}}{1+(\omega\tau)^{2}}\,\delta_{\alpha\beta},
\end{equation}
where $D_{0}=v_{F}^{2}\tau/d$ is the diffusion coefficient.
\subsection{Beyond Drude approximation.}
The diagrams that were not included in the Drude approximation of
Fig.1 contain {\it cross-correlations} between the exact electron
Green's functions $G^{R}$ and $G^{A}$ which arise because both of
them see the {\it same} random impurity potential $U({\bf r})$. We
make the simplest approximation of this potential as being the
random Gaussian field with zero mean value and zero-range
correlation function:
\begin{equation}
\label{imp} 
\langle U({\bf r})U({\bf r'})\rangle=\frac{\delta({\bf
r }-{\bf r'})}{2\pi\nu\tau}.
\end{equation}
The simplest diagrams beyond the Drude approximation are the
ladder series shown in Fig.2.
\begin{figure}[h]
%\fbox{\vtop to5cm{\vss\hsize=.7\hsize\vglue
%0cm\hspace{0.01\hsize}\hskip -.5cm \hspace{0.1\hsize}\vss}}
\includegraphics[width=110mm]{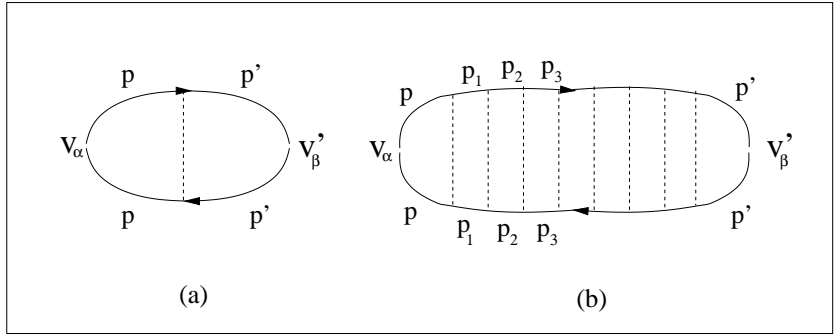}
\caption{ Ladder diagrams.} \label{Ladder}
\end{figure}
The dotted line in these diagrams represent the momentum Fourier
transform of correlation function $\langle U({\bf r})U({\bf
r'})\rangle$ which is a constant $1/2\pi\nu\tau$ independent of
the momentum transfer ${\bf p -{\bf p'}}$. This makes the
quantities $v_{\alpha}$ and $v'_{\beta}$ completely independent of
each other. Then the angular integration
\begin{equation}
\label{angular} \int d{\bf n}\, v_{\alpha}=0
\end{equation}
results in vanishing of all the ladder diagrams of Fig.2.

The next diagram with cross correlations contain an intersection
of dotted lines (see Fig.3).
\begin{figure}[h]
%\fbox{\vtop to5cm{\vss\hsize=.7\hsize\vglue
%0cm\hspace{0.01\hsize}\hskip -.5cm \hspace{0.1\hsize}\vss}}
\includegraphics[width=60mm]{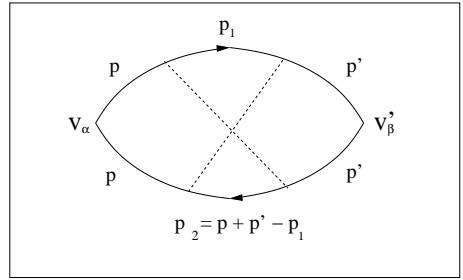}
\caption{The first of the "fan" diagrams.}
\end{figure}

In contrast to a diagram with two parallel dotted lines where all
integrations over momenta are independent, here the momentum
conservation imposes a constraint:
\begin{equation}
\label{constr} {\bf p}+{\bf p'}={\bf p_{1}}+{\bf p_{2}}.
\end{equation}
At the same time the form Eq.(\ref{GFu}) of the average Green's
function suggests that the main contribution to the momentum
integration is given by momenta confined inside a ring of the
radius $p_{F}$ and the width $1/\ell\ll p_{F}$ where $\ell$ is the
elastic mean free path $\ell=v_{F}\tau$ [see Fig.4a].
\begin{figure}[h]
%\fbox{\vtop to5cm{\vss\hsize=.9\hsize\vglue
%0cm\hspace{0.01\hsize}\hskip -.5cm \hspace{0.1\hsize}\vss}}
\includegraphics[width=110mm]{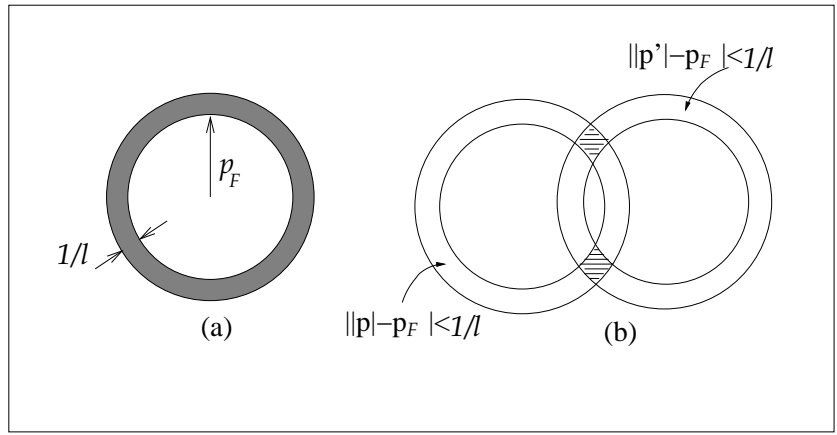}
\caption{Regions that make the main contribution to momentum
integrals.}
\end{figure}
The constraint Eq.(\ref{constr}) implies that not only ${\bf p}$
but also ${-\bf p}+{\bf p}_{1}+{\bf p}_{2}={\bf p'}$ should be
inside a narrow ring [see Fig.4b]. Thus the effective region of
${\bf p}$ integration is an {\it intersection} of two narrow rings
which volume is reduced by a large factor of $(p_{F}\ell)$
compared to an unconstrained case.
\subsection{Weak localization correction.}
The lesson one learns from the above example is that any
intersection of dotted lines in systems with dimensionality
greater than one brings about a small parameter $1/(p_{F}\ell)$.
However this does not mean that all such diagrams should be
neglected. As a matter of fact they contain an important new
physics of {\it quantum coherence} that changes completely the
behavior of low dimensional systems with $d=1,2$ at small enough
frequencies $\omega$ leading to the phenomenon of Anderson localization.

For this to happen there should be something that compensates for
the small factor $1/(p_{F}\ell)$. We will see that this is a large
return probability of a random walker or a particle randomly
scattered off impurities in low-dimensional systems. On the formal
level the corresponding diagrams are just the multiple scattering
"fan" diagrams shown in Fig.5.
\begin{figure}[h]
%\fbox{\vtop to5cm{\vss\hsize=.9\hsize\vglue
%0cm\hspace{0.01\hsize}\hskip -.5cm \hspace{0.1\hsize}\vss}}
\includegraphics[width=110mm]{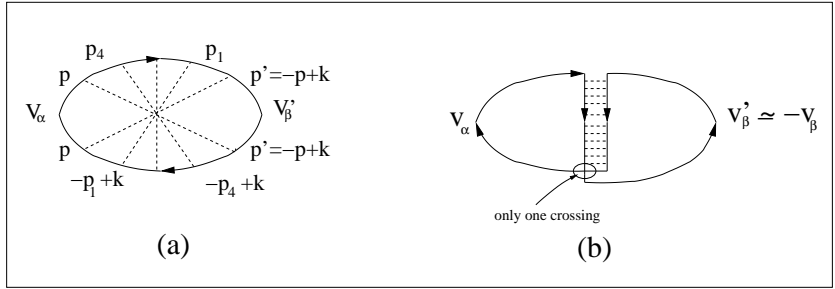}
\caption{Infinite series of "fan" diagrams.}
\end{figure}
One can see using the momentum conservation that such diagrams
have only one constraint of the type Eq.(\ref{constr})
independently of the number of dotted lines. This statement
becomes especially clear if one rewrites the fan series in a form
Fig.5b which contains a ladder series with all dotted lines
parallel to each other and only one intersection of solid lines
representing electron Green's functions. It is this ladder series
called Cooperon which describes quantum interference that leads to
Anderson localization.

To make connection between  random walks (or diffusion) in space
and a quantum correction to conductivity and to see how a
compensation of the small parameter $1/(p_{F}\ell)$ occurs we
compute the Cooperon $C({\bf k},\omega)$ for a small value of
${\bf p_{1}}+{\bf p_{2}}={\bf k}$ [see Fig.6.].
\begin{figure}[h]
%\fbox{\vtop to5cm{\vss\hsize=.9\hsize\vglue
%0cm\hspace{0.01\hsize}\hskip -.5cm \hspace{0.1\hsize}\vss}}
\includegraphics[width=110mm]{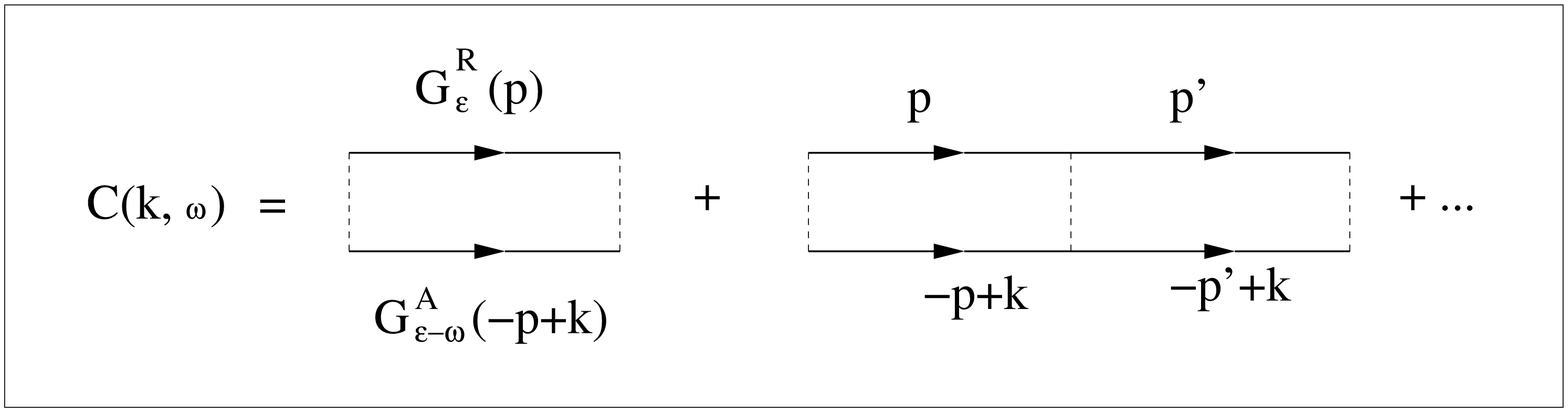}
\caption{Summation of the ladder series for a Cooperon.}
\end{figure}
This series is nothing but a geometric progression with the first
term
\begin{eqnarray}
\label{Pi-0} \Pi_{0}&=&\frac{1}{(2\pi\nu\tau)^{2}}\int d{\bf
p}\,G^{R}_{\varepsilon}({\bf p})G^{A}_{\varepsilon-\omega}(-{\bf
p}+{\bf k})
\end{eqnarray}
and the denominator $q=2\pi\nu\tau\Pi_{0}$. Thus we have
\begin{equation}
\label{Sum} C({\bf k},\omega)=\frac{\Pi_{0}}{1-q}.
\end{equation}
To compute $\Pi_{0}$ we write
\begin{eqnarray}
\label{C-RA} \Pi_{0}=\frac{\nu}{(2\pi\nu\tau)^{2}}\int d{\bf n
}\int_{-\infty}^{+\infty}\frac{d\xi}{(\varepsilon-\xi+\frac{i}{2\tau})
(\varepsilon-\omega-\xi+v_{F}{\bf n}{\bf k}-\frac{i}{2\tau})}.
\end{eqnarray}
In Eq.(\ref{C-RA})  we used an approximation $\xi_{-{\bf p}+{\bf k
}}\approx \xi_{\bf p}-v_{F}{\bf n}{\bf k}$ which is valid as long
as $|{\bf k}|\ll p_{F}$. The pole integral in Eq.(\ref{C-RA}) can
be done immediately with the result:
\begin{equation}
\label{q} q=\int d{\bf n}\,\frac{1}{1-i\omega\tau+i\ell {\bf
n}{\bf k}}.
\end{equation}
One can see a remarkable property of Eq.(\ref{q}): the denominator
of the geometric series Eq.(\ref{Sum}) is equal to 1 in the limit
$\omega\tau, |{\bf k}|\ell\rightarrow 0$. At finite but small
$\omega\tau, |{\bf k}|\ell\ll 1$ one can expand the denominator of
Eq.(\ref{q}) to obtain:
\begin{equation}
\label{diff} q\approx 1+i\omega\tau -\ell^{2}{\bf k}^{2}/d;
\;\;\;\;C({\bf k},\omega)\approx
\frac{1}{2\pi\nu\tau^{2}}\,\frac{1}{D_{0}{\bf k}^{2}-i\omega}.
\end{equation}
We immediately recognize an inverse diffusion operator in $C({\bf
k },\omega)$ which is divergent at small $\omega$ and ${\bf k}$.
This divergence is the cause of all the peculiar quantum-coherence
phenomena in systems of dimensionality $d\leq 2$.

In particular the quantum correction to conductivity given by the
diagram of Fig.5 can be written as:
\begin{equation}
\label{WAL} \delta\sigma_{\alpha \beta }=\frac{\sigma^{(D)}
}{2\pi\nu D_{0}}\,\times\frac{1}{Vol}\sum_{\bf
k}\Gamma_{\alpha\beta}\,C({\bf k},\omega).
\end{equation}
where $\Gamma_{\alpha\beta}$ is the "Hikami box" shown in Fig.7:
\begin{figure}[h]
%\fbox{\vtop to5cm{\vss\hsize=.9\hsize\vglue
%0cm\hspace{0.01\hsize}\hskip -.5cm \hspace{0.1\hsize}\vss}}
\includegraphics[width=110mm]{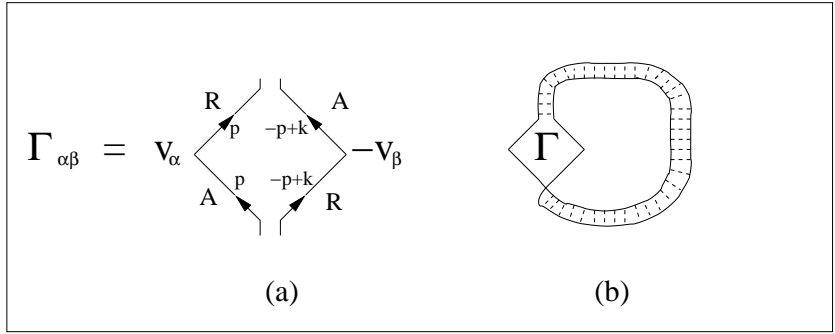}
\caption{The diagram representation for the Hikami box (a) and for
the quantum correction to conductivity (b).} \label{Hikami}
\end{figure}
Its analytic expression is given by:
\begin{equation}
\label{HB} \int d{\bf p}\,{\bf
v}_{\alpha}({\bf p}){\bf v}_{\beta}(-{\bf p}+{\bf k})\,
G^{R}_{\varepsilon}({\bf p})G^{A}_{\varepsilon-\omega}({\bf p})
G^{R}_{\varepsilon}(-{\bf p}+{\bf
k})G^{A}_{\varepsilon-\omega}(-{\bf p}+{\bf k}).
\end{equation}
In the limit $\omega\tau\ll 1$ and $|{\bf k}|\ell\ll 1$ one can
set $\omega={\bf k}=0$ in Eq.(\ref{HB}). Then ${\bf v}(-{\bf
p}+{\bf k})= -{\bf v}(\bf p)=- v_{F}{\bf n}$ and after the pole
integration over $\xi$ and the angular integration over ${\bf n}$
we  obtain for $\Gamma_{\alpha\beta}$:
\begin{equation}
\label{HBres} -\int d{\bf n}\,v_{F}^{2}{\bf
n}_{\alpha}{\bf n}_{\beta}\int_{-\infty}^{+\infty}\frac{\nu\,d\xi}
{(\varepsilon-\xi+\frac{i}{2\tau})^{2}(\varepsilon-\xi-\frac{i}{2\tau})^{2}}
=-4\pi\nu\tau^{2}D_{0}.
\end{equation}
Finally using Eq.(\ref{WAL}) we get:
\begin{equation}
\label{WL} \sigma(\omega) =
\sigma^{(D)}\,\left(1-\frac{1}{\pi\nu}\,\frac{1}{Vol}\Re\sum_{\bf
k}\frac{1}{D_{0}{\bf k}^{2}-i\omega} \right).
\end{equation}
This is the celebrated formula for weak Anderson localization. One
can see that the quantum correction is given by the sum over
momenta of the diffusion propagator, that is, it is proportional
to the {\it return probability} at a time $\sim 1/\omega$ for a
random walker in the $d$-dimensional space. A remarkable property
of random walks in low-dimensional space is that the return
probability increases with time. That is why the  quantum
correction to conductivity increases with decreasing the frequency
$\omega$ as $1/\sqrt{\omega}$ in a quasi-one dimensional wire and
as $\log(1/\omega)$ in a two-dimensional disordered metal.

The structure of Eq.(\ref{WL}) suggests  a qualitative picture of
weak Anderson localization. It is an interference of two
trajectories with a loop that differ only in the direction of
traversing the loop [Fig.8].
\begin{figure}[h]
%\fbox{\vtop to7cm{\vss\hsize=.7\hsize\vglue
%0cm\hspace{0.01\hsize}\hskip -.5cm \hspace{0.1\hsize}\vss}}
\includegraphics[width=110mm]{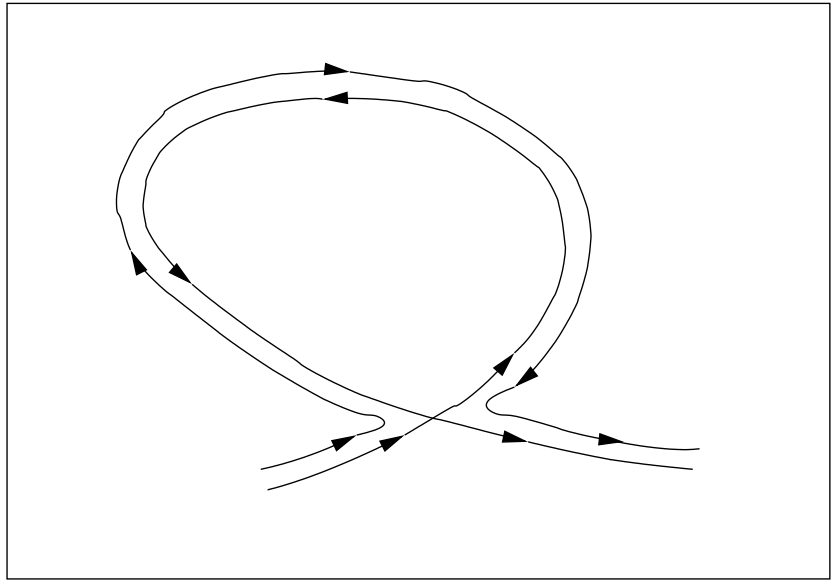}
\caption{Two interfering trajectories with loops} \label{loop}
\end{figure}
Although the phase that corresponds to each trajectory is large
and random, the phase {\it difference} between them is zero
because of the time-reversal invariance. As a result they
can interfere  and amount to a creation of a {\it random
standing wave } which is the paradigm of Anderson localization.

\section{Non-linear response to a time-dependent perturbation}
In this section we review the main steps of the Keldysh formalism
\cite {Keld, Ram-Smth} which are necessary to describe a 
quantum system of
non-interacting electrons subject to external time-dependent
perturbation $\hat{V}(t)$.
\subsection{General structure of nonlinear response function.}
The matrix Keldysh Green's function
\begin{equation}
\label{KGF} {\bf G}=\left(\matrix{G^{R}&G^{K}\cr 0&
G^{A}\cr}\right)
\end{equation}
contains -- besides familiar retarded and advanced Green's
functions ${\bf G}^{11}=G^{R}$ and ${\bf G}^{22}=G^{A}$ -- also
the third, Keldysh function ${\bf G}^{12}=G^{K}$. The latter is
the only one needed to compute an expectation value of any
operator $\hat{O}$ both in equilibrium and beyond:
\begin{equation}
\label{ob} O=-i Tr(\hat{O}G^{K}).
\end{equation}
The retarded and advanced Green's functions appear only at the
intermediate stage to make it possible to write down Dyson
equations in the matrix form. It is of principal importance that
the component ${\bf G}^{21}$ is zero. One can show that this is a
consequence of causality \cite{Kanzieper}. In equilibrium there is a 
relationship
between $G^{K}$ and $G^{R/A}$. It reads:
\begin{equation}
\label{K-RA}
G_{\varepsilon}^{K}=(G^{R}-G^{A})_{\varepsilon}\,\tanh\left(\frac{\varepsilon}{2T}
\right).
\end{equation}
As without time-dependent perturbation the system is in
equilibrium, the corresponding components
$G_{0,\varepsilon}^{R/A}$ and $G_{0,\varepsilon}^{K}$ of the
matrix Keldysh Green's function obey the relationship
Eq.(\ref{K-RA}).

In the presence of the perturbation one can write the Dyson
equation:
\begin{equation}
\label{DyE} {\bf G}={\bf G}_{0}+{\bf G}_{0}\hat{{\bf V}}{\bf G},
\end{equation}
where the operator of {\it external} time-dependent perturbation
$\hat{{\bf V}}(t)$ is proportional to a unit matrix in the Keldysh
space.
\begin{figure}[h]
%\fbox{\vtop to5cm{\vss\hsize=.9\hsize\vglue
%0cm\hspace{0.01\hsize}\hskip -.5cm \hspace{0.1\hsize}\vss}}
\includegraphics[width=110mm]{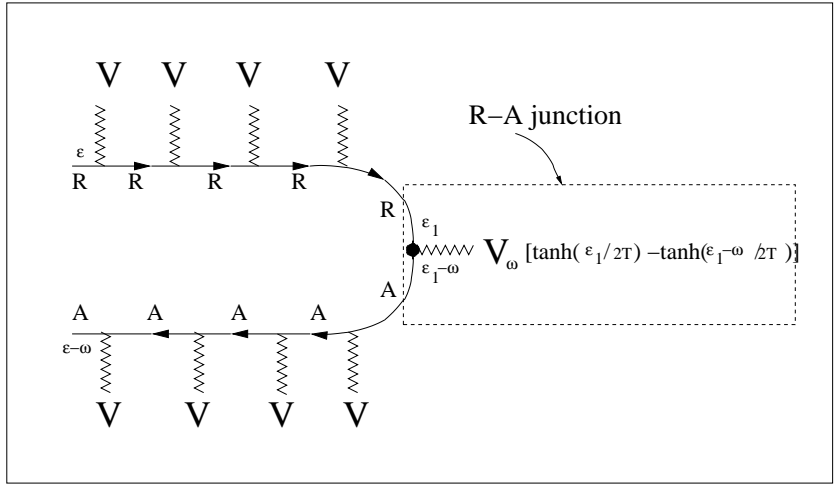}
\caption{The diagram representation of the "anomalous" term in the
Keldysh function.} \label{G-a}
\end{figure}
The structure of perturbation series for $G^{K}$ that corresponds
to Eq.(\ref{DyE}) is as follows:
\begin{equation}
\label{str}
G^{K}=G_{0}^{12}+G_{0}^{11}\hat{V}G_{0}^{12}+G_{0}^{12}\hat{V}G_{0}^{22}+
G_{0}^{11}\hat{V}...G_{0}^{11}\hat{V}G_{0}^{12}\hat{V}G_{0}^{22}...G_{0}^{22}+...
\end{equation}
As $G_{0}^{21}=0$ and
$G_{0}^{12}=(G^{R}_{0}-G^{A}_{0})\,\tanh(\varepsilon/2T)$  each term of
the perturbation series Eq.(\ref{str}) is a string of successive
$G^{R}_{0}$ functions followed by a string of $G^{A}_{0}$ functions with
only one switching point between them where the factor
$\tanh(\varepsilon/2T)$ is attached to. For the same reason the
components $G^{R}$ and $G^{A}$ are the series that contain only
$G^{R}_{0}$ or $G^{A}_{0}$, respectively:
\begin{equation}
\label{str-RA} G^{R/A}=G_{0}^{R/A}+G_{0}^{R/A}\hat{V}G_{0}^{R/A}+
G_{0}^{R/A}\hat{V}G_{0}^{R/A}\hat{V}G_{0}^{R/A}+...
\end{equation}
Using Eq.(\ref{str-RA}) one can replace the string of
$G_{0}^{R/A}$ functions in Eq.(\ref{str}) by the exact retarded or
advanced Green's function $G^{R/A}_{\varepsilon, \varepsilon'}$
that depends on two energy variables because of the breaking of
translational invariance in the time domain by a time dependence
of perturbation. The result is:
\begin{eqnarray}
\label{K-anom}
& &G^{K}_{\varepsilon,\varepsilon'}=G^{R}_{\varepsilon,\varepsilon'}\,
\tanh\left
(\frac{\varepsilon'}{2T}\right)-\tanh\left
(\frac{\varepsilon}{2T}\right)\,G^{A}_{\varepsilon,\varepsilon'}+
\\ \nonumber &+&\int 
\frac{d\varepsilon_{1}}{2\pi}\int\frac{d\omega}{2\pi}
G^{R}_{\varepsilon,\varepsilon_{1}}\hat{V}(\omega)G^{A}_{\varepsilon_{1}-
\omega,
\varepsilon'}\,\left[\tanh\left
(\frac{\varepsilon_{1}}{2T}\right)-\tanh\left
(\frac{\varepsilon_{1}-\omega}{2T}\right)\right].
\end{eqnarray}
The first two terms in Eq.(\ref{K-anom}) just reproduce the
structure of the equilibrium Keldysh function Eq.(\ref{K-RA}). The
most important for us will be the last, so called "anomalous" term
$G^{{\rm anom}}$ in Eq.(\ref{K-anom}) as it contains both
$G^{R}_{0}$ and $G^{A}_{0}$ functions which makes nonzero the pole
integrals over $\xi$ and allows to build a Cooperon. This term is
graphically represented in Fig.9.

One can see that of all the vertices with a perturbation operator
$\hat{V}$ one is special: it is a switching point between the
strings of retarded and advanced Green's functions, so called
$R-A$ junction \cite{Kanzieper}. It is convenient to switch from the 
energy to the
time domain. In this representation the anomalous part of $G^{K}$
reads \cite{Kanzieper}:
\begin{eqnarray}
\label{anom-t} G^{{\rm anom}}_{t,t'}=
G^{R}_{t,t_{1}}\,G^{A}_{t_{2},t'}\,
(\hat{V}(t_{2})-\hat{V}(t_{1}))\,h_{0}(t_{1}-t_{2}).
\end{eqnarray}
In Eq.(\ref{anom-t}) we denoted the Fourier transform of
$\tanh(\varepsilon/2T)$ as:
\begin{equation}
\label{h} h_{0}(t)=\int \frac{d\varepsilon}{2\pi}\,
e^{i\varepsilon t}\,\tanh(\varepsilon/2T)=\frac{i T}{\sinh(\pi T
t)}.
\end{equation}
We also assume integration over repeated time variables.

The graphic representation of Eq.(\ref{anom-t}) is given in
Fig.10.
\begin{figure}[h]
%\fbox{\vtop to5cm{\vss\hsize=.9\hsize\vglue
%0cm\hspace{0.01\hsize}\hskip -.5cm \hspace{0.1\hsize}\vss}}
\includegraphics[width=100mm]{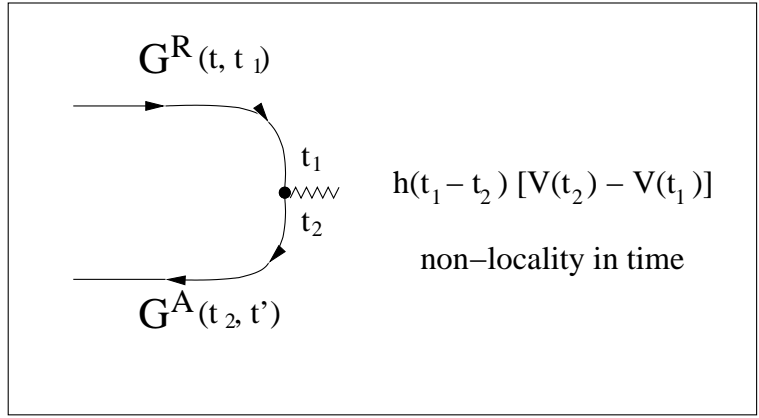}
\caption{Anomalous contribution to $G^{K}$ in the time domain.}
\label{t-nonloc}
\end{figure}
Eqs.(\ref{K-anom}),(\ref{anom-t}) give a general structure of
nonlinear response to a time-dependent perturbation. The main
feature is the distinction between a regular $\hat{V}$ vertex that
connects two Green's functions with the same analytical properties
(both $G^{R}$ or both $G^{A}$) and the $R-A$ junction. One can see
that the time variables do not match at the R-A junction because
of the attached energy distribution function, while in at any
regular vertex $\hat{V}$ they do match.
\subsection{Approximation of single photon absorption/emission}
While an exact solution of the nonlinear response is a difficult
task, there is an essentially nonlinear regime where a regular
solution can be found. The corresponding approximation consists of
expanding each {\it disorder averaged} retarded or advanced
Green's function up to the second order in $\hat{V}$:
\begin{equation}
\label{approxi} \langle G^{R/A}\rangle\approx \langle
G_{0}^{R/A}\rangle +\langle G_{0}^{R/A}\rangle\hat{V}\langle
G_{0}^{R/A}\rangle+\langle G_{0}^{R/A}\rangle\hat{V} \langle
G_{0}^{R/A}\rangle \hat{V}\langle G_{0}^{R/A}\rangle
\end{equation}
Yet, since the ladder series (a Cooperon) contains an {\it
infinite} number of retarded (advanced) Green's functions, this
approximation leads to an essentially non-linear results that do
not reduce to a finite order response function. We will see that
this approximation implies a {\it sequential} absorption/emission
of photons rather than the {\it multiple-photon} processes.

Another approximation which is similar to $\omega\tau\ll 1$ is
based on the fact that the disorder average Green's function
Eq.(\ref{GFu}) decays exponentially in the time domain:
\begin{equation}
\label{tau-dec} \langle G_{0}^{R/A}\rangle_{t}=\mp i \,\theta(\pm
t)\,e^{-i t\xi_{{\bf p}}}\,e^{-t/\tau}.
\end{equation}
Assuming that the momentum relaxation time is much shorter than
the characteristic scale (e.g. the period) of a time-dependent
perturbation one can approximate $\langle G_{0}^{R/A}\rangle_{t}$
by a $\delta$-function and its derivative:
\begin{equation}
\label{deltaf} \langle G_{0}^{R/A}\rangle_{t}\approx \delta(t)\,
\langle G_{0}^{R/A}\rangle_{\varepsilon=0}-i\partial_{t}\delta(t)
\langle G_{0}^{R/A}\rangle_{\varepsilon=0}^{2}.
\end{equation}
\section{Quantum rectification by a mesoscopic ring.}
To proceed further on we need to make some assumptions about the
time-dependent perturbation and specify an observable of interest. 

In this section we consider a problem of rectification of an  {\it ac 
signal}
by a disordered metal. It has first been considered in Ref.\cite{Falko}
in a single-connected geometry where the entire effect is due to 
mesoscopic fluctuations. Here we focuse on the case of 
a quasi-one dimensional disordered metal ring pierced
by a magnetic flux $\phi(t)$ that contains both a constant part
$\phi$ and an oscillating part $\phi_{ac}(t)$ \cite{KraYud,ArKra,KrAlt}. 
In this case the topology of a ring and the presence of a constant
magnetic flux makes it meaningful to study the {\it disorder-average}
rectification effect.

The time-dependent perturbation in
this case is:
\begin{equation}
\label{J}
\hat{V}=-\hat{v}_{x}\,\varphi(t),\;\;\;\;\varphi(t)=\frac{2\pi}{L}\,
\frac{\phi_{ac}(t)}{\phi_{0}},
\end{equation}
where $L$ is the circumference of the ring, and $\phi_{0}=hc/e$ is
the flux quantum.  We assume that the ring curvature is large
compared to all microscopic lengths in the problem so that it can
be replaced by quasi-one dimensional wire along the $x$-axis with
the twisted boundary conditions $\Psi(L)=\Psi(0)\;\exp[2\pi i 
\phi/\phi_{0}]$.

To solve this problem we need to take into account the ac
perturbation only in the denominator $q$ of the geometric series
Eq.(\ref{Sum})that determines the Cooperon. This is because $1-q$
is governed by {\it small corrections} with the energy scale much
smaller than $1/\tau$. The corrections due to the ac flux
perturbation are given by the diagrams in Fig.11.
\begin{figure}[h]
%\fbox{\vtop to7cm{\vss\hsize=.9\hsize\vglue
%0cm\hspace{0.01\hsize}\hskip -.5cm \hspace{0.1\hsize}\vss}}
\includegraphics[width=89mm]{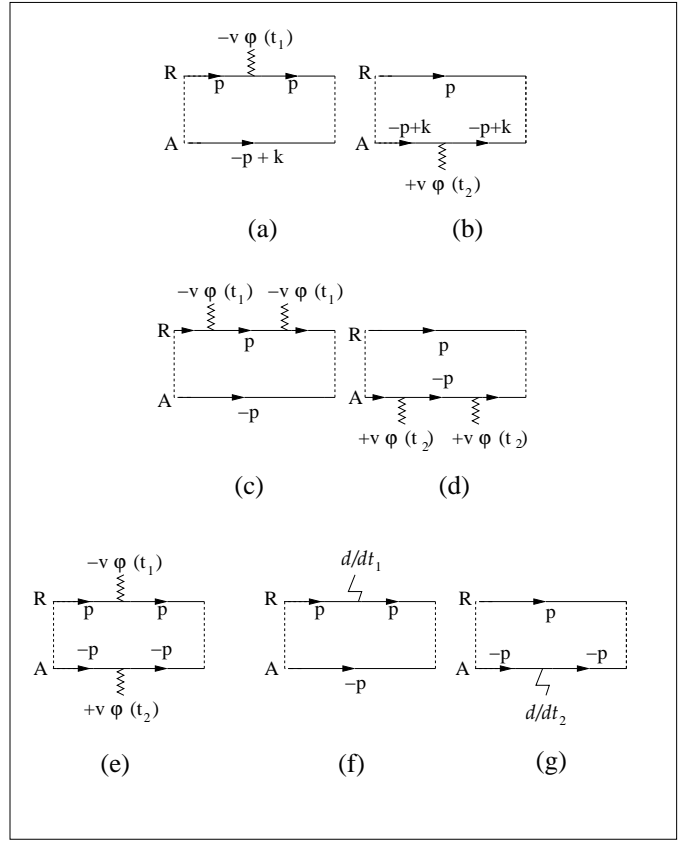}
\caption{The ac flux and time-derivative corrections to the inverse 
Cooperon operator.} \label{ac corr}
\end{figure}
Note that in linear in $\hat{V}$ corrections one should take into
account the small momentum ${\bf k}$ whereas in quadratic in
$\hat{V}$ corrections ${\bf k}$ can be set zero. All the
corrections of Fig.11 can be computed by doing the pole integrals
over $\xi$ and angular integrals over ${\bf n}$ as in the previous
section. Note also that $1-q$ is essentially the {\it inverse}
Cooperon operator. For an arbitrary time dependence of
$\varphi(t)$ one should replace $-i\omega$ in Eq.(\ref{diff}) by
the time derivative $\partial_{t_{1}}-\partial_{t_{2}}$ which one
can obtain from the second term in Eq.(\ref{deltaf}). Such a
structure of the time derivatives implies that the sum of time
arguments $t_{1}+t_{2}=t_{1}'+t_{2}'$ is conserved [see Fig.12]
which is the consequence of the constant density of states
approximation.
 \begin{figure}[h]
%\fbox{\vtop to4cm{\vss\hsize=.9\hsize\vglue
%0cm\hspace{0.01\hsize}\hskip -.5cm \hspace{0.1\hsize}\vss}}
\includegraphics[width=90mm]{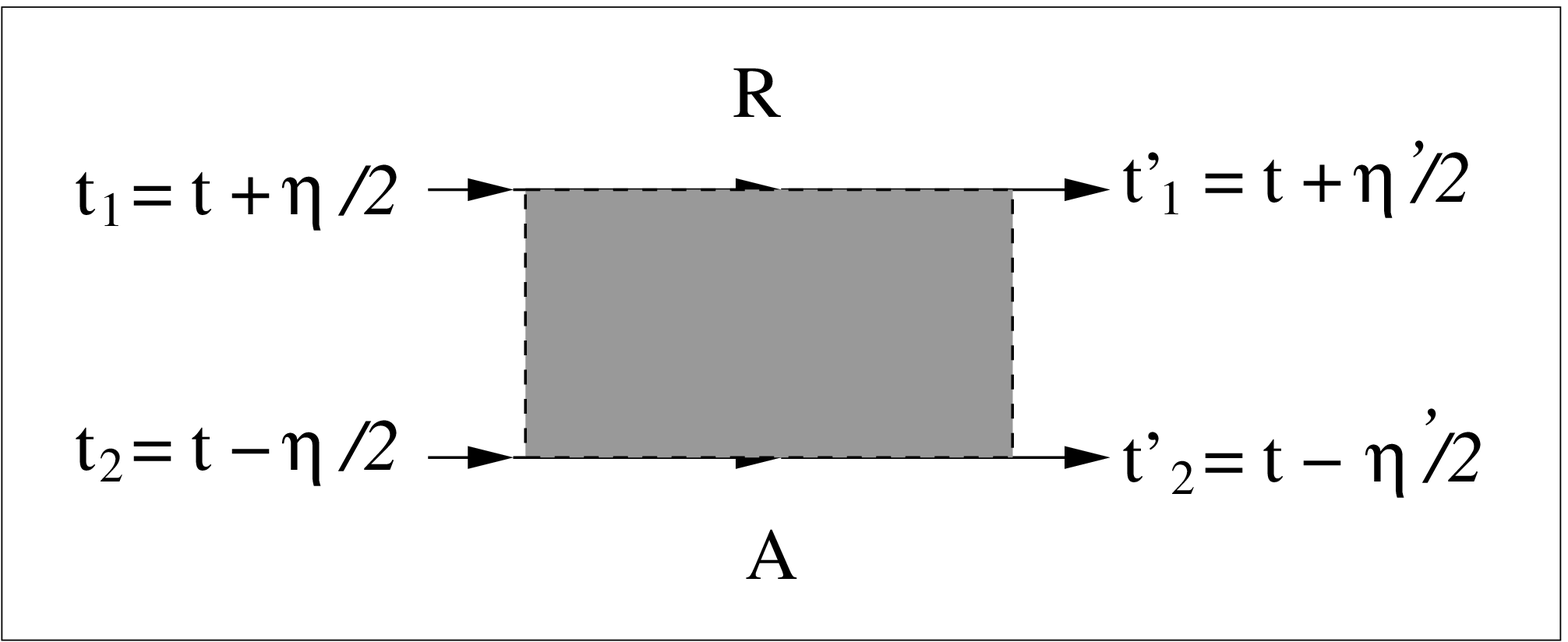}
\caption{The time arguments in the Cooperon.} \label{t-Coop}
\end{figure}
As a result, the equation for a time-dependent Cooperon
takes the form \cite{AAKh, Vavilov, Kanzieper}:
\begin{equation}
\label{time-Coop}
\left\{\frac{\partial}{\partial\eta}+\frac{D_{0}}{2}\left[\varphi
(t+\eta/2) +\varphi(t-\eta/2)-{\bf k}_{x}\right]^{2}
\right\}\,C_{t}(\eta,\eta';{\bf k}_{x})=
\frac{\delta(\eta-\eta')}{2\pi\nu\tau^2},
\end{equation}
where the momentum ${\bf 
k}_{x}(\phi)=k_{m}(\phi)=(2\pi/L)(m-2\phi/\phi_{0})$
is quantized according to the twisted boundary conditions.

For completeness we give also an equation for the time-dependent
{\it diffuson} [Fig.13]:
\begin{equation}
\label{time-Diff} \left\{\frac{\partial}{\partial
t}+D_{0}\left[\varphi (t+\eta/2) -\varphi(t-\eta/2)-{\bf
k}_{x}\right]^{2} \right\}\,D_{\eta}(t,t';{\bf k}_{x})=
\frac{\delta(t-t')}{2\pi\nu\tau^2},
\end{equation}
where ${\bf k}_{x}=(2\pi/L)\,m$ is independent of the DC flux $\phi$.

In this case the structure of the time-derivative
$\partial_{t_{1}}+\partial_{ t_{2} }$ suggests that the {\it
difference} of the time arguments is conserved
$t_{1}-t_{2}=t_{1}'-t_{2}'$.
 \begin{figure}[h]
%\fbox{\vtop to4cm{\vss\hsize=.9\hsize\vglue
%0cm\hspace{0.01\hsize}\hskip -.5cm \hspace{0.1\hsize}\vss}}
\includegraphics[width=90mm]{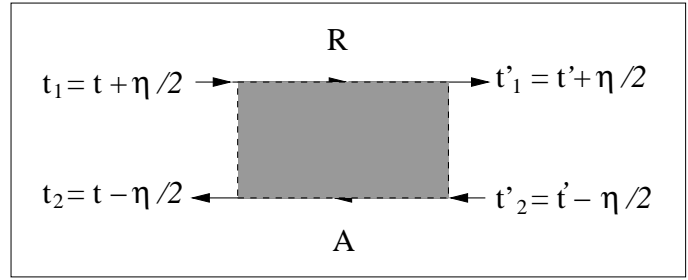}
\caption{The time-dependent Diffuson.} \label{t-Diff}
\end{figure}

In general the current response $I(t)$ to the electric field
$E(t-\tau)$ is given by:
\begin{equation}
\label{CR} I(t)=\int_{0}^{\infty}K(t,\tau)\,E(t-\tau)\,d\tau,
\end{equation}
where in our particular case $E(t)=-\partial_{t}\varphi(t)$.

At small $1/p_{F}\ell$ the main contribution to the {\it
nonlinear} response function is that of the quantum coherence
correction shown in Figs.5,7. All what we have to do in order to
compute this nonlinear response function is to substitute the
time-dependent Cooperon into Eq.(\ref{WAL}) and take care of all
the time arguments [see Fig.14]. Namely, (i) according to 
Eq.(\ref{deltaf}) 
the 
time arguments
corresponding to the beginning and the end of any solid line
representing the disorder averaged functions $\langle G_{0}^{R/A}
\rangle$ should be the same and (ii) the sum (difference)of
"incoming times" in the Cooperon (Diffuson)  are equal to the sum
(difference)of "outgoing times" as in Figs.12,13. 
 \begin{figure}[h]
%\fbox{\vtop to5cm{\vss\hsize=.9\hsize\vglue
%0cm\hspace{0.01\hsize}\hskip -.5cm \hspace{0.1\hsize}\vss}}
\includegraphics[width=110mm]{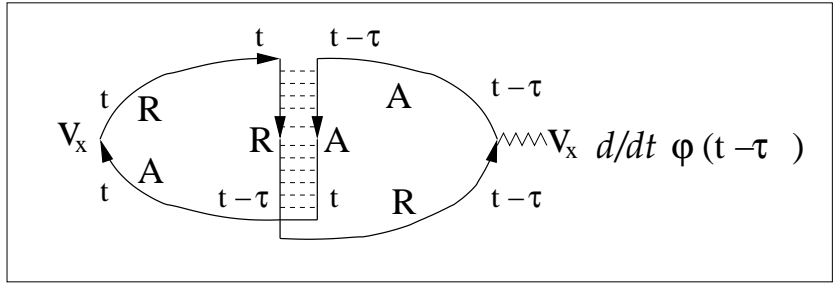}
\caption{Nonlinear current response function and the Cooperon.}
\label{response}
\end{figure}
Note that application of rules (i),(ii) to the diagram of Fig.14 leads to 
the coinciding time arguments 
$t_{1}\rightarrow t_{2}=t-\tau$ in the corresponding retarded-advanced 
junction 
(see Fig.10). This ensures fulfillment of Eq.(\ref{CR}), as 
$h_{0}(t_{1}-t_{2})\,(V(t_{2})-V(t_{1}))\rightarrow 
v_{x}\,\partial_{t}\varphi(t-\tau)\propto E(t-\tau)$.
 
The disorder average current response function
$K(t,\tau)$ is equal  to:
\begin{equation}
\label{cresp} K(t,\tau)=\sigma^{D} \tau^{-1}e^{-t/\tau}
-\frac{4e^{2}D_{0}}{h}\,\tilde{C}_{t-\tau/2}(\tau,-\tau),
\end{equation}
where $\tilde{C}_{t}(\eta,\eta')=(2\pi\nu\tau^{2}/Vol)\,\sum_{{\bf
k}_{x}}C_{t}(\eta,\eta';{\bf k}_{x})$ and $C_{t}(\eta,\eta';{\bf
k}_{x})$ is the solution to Eq.(\ref{time-Coop}):
\begin{eqnarray}
\label{sol} & & 2\pi\nu\tau^{2}C_{t}(\eta,\eta';{\bf
k}_{x})=\theta(\eta-\eta')\times\\ \nonumber & &
\exp\left\{-\frac{D_{0}}{2}\int_{\eta'}^{\eta}\left[
\varphi(t+\zeta/2)+\varphi(t-\zeta/2)-{\bf
k_{x}}\right]^{2}\,d\zeta \right\}.
\end{eqnarray}
The first term in Eq.(\ref{cresp}) is just the linear response given by 
the diagrams Fig.1c,d.
The second term corresponds to the quantum-coherent contribution 
of Fig.14. The fact that
$\tilde{C}_{t-\tau/2}(\tau,-\tau)$ depends not only on $\tau$ but
also on $t$ makes it possible to have  dc response caused by ac
electric field $E(t)=-\partial_{t}\varphi(t)$.

For a particular case of a ring with a constant and an {\it ac} magnetic
flux we note that the {\it dc} response arises only from {\it odd}
terms of expansion of Eq.(\ref{sol}) in powers of the ac
perturbation $\varphi(t)$. As odd terms in $\varphi(t)$ enter in a 
combination $\varphi(t){\bf k}_{x}$ the {\it dc} response involves only 
{\it odd} terms in ${\bf k}_{x}=k_{m}=(2\pi/L)(m-2\phi/\phi_{0})$. If
the constant magnetic flux  $\phi=0$ or we neglect quantization of
momentum and do an integral over ${\bf k}_{x}$ instead of a sum, the 
result for the odd in $\varphi(t)$ part of
$\tilde{C}_{t}(\eta,\eta')$ is zero. Otherwise it is {\it
periodic} in the flux $\phi$ with the period $\phi_{0}/2$, since
$k_{m}(\phi+\phi_{0}/2)=k_{m-1}(\phi)$ and the summation is over
all integer $m$. We see that the dc response to the ac perturbation, or
the rectification of the ac flux by an ensemble of mesoscopic
rings, is an essentially quantum, Bohm-Aharonov-like effect.
Furthermore, it is an {\it odd} in the flux Bohm-Aharonov effect
that can be represented by the Fourier series:
\begin{equation}
\label{Fou} \langle I_{dc}(\phi)\rangle=\sum_{n=1}^{\infty}
I_{n}\,\sin\left(\frac{4\pi n\phi}{\phi_{0}} \right).
\end{equation}
Expression Eq.(\ref{Fou}) has the same symmetry and periodicity in the 
magnetic flux $\phi$ as the disorder-averaged eqililibrium persistent 
current \cite{ButImLan, Lev}. However, its magnitude may be much larger 
(in the grand-canonical ensemble considered here the 
disorder-averaged persistent current is strictly zero). 

Applying the Poisson summation trick
\begin{eqnarray}
\label{Psum} &&\sum_{m}f(m-\phi)=\int dx\,
f(x-\phi)\sum_{m}\delta(x-m)=\\ \nonumber &=&\int dx\, f(x-\phi)\,
\sum_{n}e^{2\pi i n x}=\sum_{n}e^{2\pi i n \phi}\,\int dx\,
e^{2\pi i n x}\,f(x)
\end{eqnarray}
 to Eq.(\ref{sol}) one obtains
\begin{equation}
\label{I-n} I_{n}= \frac{4i e D_{0}}{\pi
L}\,\int_{0}^{\infty}d\tau\,\overline{
C^{(n)}_{t}(\tau)\partial_{t}\varphi(t-\tau/2)},
\end{equation}
where the overline means averaging over time $t$ and
\begin{equation}
\label{CC}
C_{t}^{(n)}(\tau)=\sqrt{\frac{\tau_{D}}{4\pi\tau}}\,
e^{-\frac{n^{2}\tau_{D}}{4\tau}}\,
e^{i n S_{1}[\varphi]}\,e^{-\tau S_{2}[\varphi]}.
\end{equation}
Here $\tau_{D}=L^{2}/D_{0}$ is time it takes for a diffusing
particle to go around the ring, and $S_{1,2}$ are defined as
follows:
\begin{equation}
\label{S-1} S_{1}[\varphi]=2
L\,\left[\frac{1}{\tau}\int_{t-\tau/2}^{t+\tau/2}\varphi(t_{1})\,dt_{1}
\right]\equiv 2 L\,\langle \varphi_{t_{1}}\rangle_{t;\tau},
\end{equation}
\begin{equation}
\label{S-2} S_{2}[\varphi]=2D_{0}\, \left[\langle
\varphi_{t_{1}}^{2}\rangle_{t;\tau}+\langle
\varphi_{t_{1}}\varphi_{2t-t_{1}}\rangle_{t;\tau}-2\langle
\varphi_{t_{1}}\rangle_{t;\tau}^{2} \right].
\end{equation}
Eqs.(\ref{CC})-(\ref{S-2}) are valid for an arbitrary
time-dependence of $\varphi(t)$. However, they take the simplest
form for a noise-like ac flux with the small  correlation time
$\tau_{0}\ll \tau_{D}$ (but we assume $\tau_{0}\gg \tau$). In this case 
the
second and the third terms in Eq.(\ref{S-2}) can be neglected and
the first term reduces to a {\it constant} that determines the
{\it dephasing time} caused by ac noise:
\begin{equation}
\label{tau}
\frac{1}{\tau_{\varphi}}=2D_{0}\overline{\varphi^{2}(t)}.
\end{equation}
Then the time averaging in Eq.(\ref{I-n}) reduces to
\begin{equation}
\label{time-av}
-i\overline{\partial_{t}\varphi(t-\tau/2)\exp\left\{
\frac{in}{\tau}\,2L\,\int_{t-\tau/2}^{t+\tau/2}\varphi(t_{1})\,dt_{1}\right\}}
=\frac{ n}{\tau}\,2L\,\overline{\varphi^{2}(t)}.
\end{equation}
Since the time-average of the a total time-derivative is zero, we
can transfer the differentiation to the exponent. For the case of
a white-noise ac flux only the lower limit of integration in the
exponent should be differentiated, the quantity $\langle
\varphi_{t_{1}}\rangle_{t;\tau}$ being set zero afterwards. The
remaining integral over $\tau$ is done exactly:
\begin{equation}
\label{integ}
\int_{0}^{\infty}\frac{d\tau}{\tau^{3/2}}\,\exp\left[-\frac{n^{2}\tau_{D}}
{4\tau}-\frac{\tau}{\tau_{\varphi}}
\right]=\frac{2\sqrt{\pi}}{n\,\sqrt{\tau_{D}}}\,
e^{-n\sqrt{\tau_{D}/\tau_{\varphi}}}
\end{equation}
 \begin{figure}[h]
%\fbox{\vtop to7cm{\vss\hsize=.7\hsize\vglue
%0cm\hspace{0.01\hsize}\hskip -.5cm \hspace{0.1\hsize}\vss}}
\includegraphics[width=100mm]{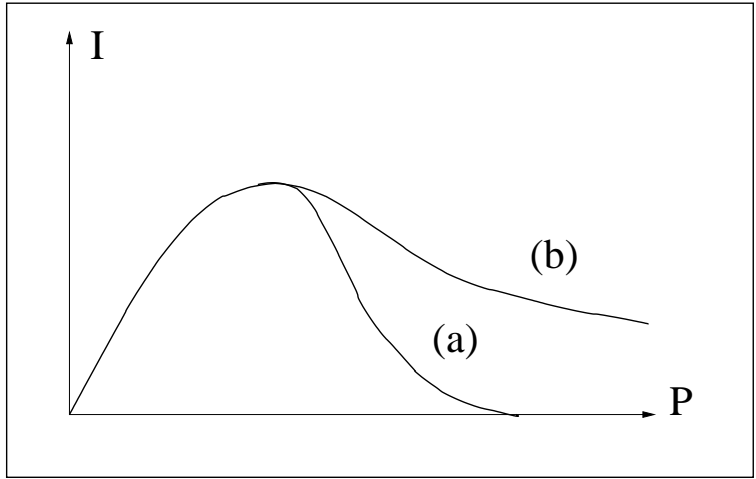}
\caption{The dependence of $I_{1}$ on the ac power 
$P=\overline{\varphi^{2}(t)}$ for a
white-noise (a) and harmonic (b) perturbation.}
\label{current-power}
\end{figure}
Finally we arrive at a remarkably simple result for the
disorder-average dc current generated in mesoscopic rings by a
white-noise ac perturbation \cite{KrAlt}:
\begin{equation}
\label{AKr} I_{n}=-\frac{4}{\pi}\,
\left(\frac{e}{\tau_{\varphi}}\right)\,\exp\left[-n\frac{L}{L_{\varphi}}
\right],
\end{equation}
where the {\it dephasing length} is equal to:
\begin{equation}
\label{L-fi}
\frac{1}{L_{\varphi}^{2}}=\frac{1}{D_{0}\tau_{\varphi}}=
2\overline{\varphi^{2}(t)}
\end{equation}
It follows from Eq.(\ref{AKr}) that the weak ac white noise produces a 
net dc current
in an ensemble  of mesoscopic rings which is of the order of
$e/\tau_{\varphi}$ where $\tau_{\varphi}$ is the dephasing time
produced by the same ac noise. At an ac power large enough to
produce a dephasing length smaller than the circumference of a
ring times the winding number $n$, the destructive effect of ac
perturbation prevails and the current $I_{n}$ decreases
exponentially [see Fig.15].

Note that the tail of the dependence $I_{n}$ on the ac power is
very sensitive to the correlations in the ac perturbation at
different times. For instance, in the case of harmonic
perturbation where $\varphi(t)$ has an infinite range
time-correlations, the current decreases very slowly, only as the
inverse square-root of the ac power \cite{KraYud}. We will see later on 
that
this is related with the phenomenon of {\it no-dephasing points}.
\section{Diffusion in the energy space.}
In this section we apply the Keldysh formalism outlined above to the 
problem
of {\it heating} by external time-dependent perturbation. The main
object to study will be the {\it non-equilibrium} electron energy
distribution function:
\begin{equation}
\label{f-e} f(\varepsilon;t)=\frac{1}{2}-\frac{1}{2}\int d\eta\,
\,h_{t}(\eta )\,e^{-i\varepsilon\eta}
\end{equation}
which is related to the Keldysh function $G^{K}(t,t')\equiv 
Vol^{-1}\int d^{d}{\bf r}\; G^{K}(t,t';{\bf r},{\bf r})$ at coincident 
space variables (averaged over volume): 
\begin{equation}
\label{f-K} G^{K}(t+\eta/2,t-\eta/2)=-2\pi i\nu\,h_{t}(\eta).
\end{equation}
Eqs.(\ref{f-e}),(\ref{f-K}) generalize Eq.(\ref{K-RA}) to the case
of non-equilibrium, time-dependent energy distribution function.
The total energy ${\cal E}(t)$ of electron system can be expressed
in terms of $f(\varepsilon,t)$:
\begin{equation}
\label{energy} {\cal E}(t)=\nu\;Vol \int
d\varepsilon\,\varepsilon\,[f(\varepsilon,t)-\theta(-\varepsilon)]+const.
\end{equation}
Then the time-dependent {\it absorption rate}
$W(t)=\partial_{t}{\cal E}(t)$ is given by:
\begin{equation}
\label{W}
 W(t)=-\frac{Vol}{2}\,\lim_{\eta\rightarrow
0}\,\partial_{t}\,\partial_{\eta}\,
G^{K}\left(t+\frac{\eta}{2},t-\frac{\eta}{2}
\right)=\frac{i\pi}{\delta}\,\,\lim_{\eta\rightarrow
0}\,\partial_{t}\,\partial_{\eta}\,h_{t}(\eta),
\end{equation}
where $\delta=(\nu\, Vol)^{-1}$ is the mean separation between
electron energy levels (mean level spacing).

For tutorial reasons in this section we consider a specific model
system described by the Hamiltonian:
\begin{equation}
\label{Ham} \hat{H}=\varepsilon(\hat{{\bf p}})+U({\bf r})+V({\bf r})
\,\varphi(t),
\end{equation}
where not only $U({\bf r})$ given by Eq.(\ref{imp}) but also
the perturbation potential $V({\bf r})$ is a Gaussian random field
which is statistically independent of $U({\bf r})$ and is
described by the correlation function:
\begin{equation}
\label{Gamma} \langle V({\bf r})V({\bf r'})
\rangle=\frac{\Gamma}{\pi\nu}\,\delta({\bf r}-{\bf r'}).
\end{equation}
This model is the simplest example of a {\it potential ac source}
in contrast to the {\it flux ac source} considered in the previous
section. In the low-frequency $\omega\tau_{D}\ll 1$ and the
modestly low ac intensity $\Gamma\tau\ll 1$ limits this model is
equivalent to the {\it random matrix theory} with the
time-dependent Hamiltonian \cite{Kanzieper, Vavilov}:
\begin{equation}
\label{RMT} \hat{H}_{RMT}=\hat{H}_{0}+\hat{V}\,\varphi(t),
\end{equation}
where both $\hat{H}_{0}$ and $\hat{V}$ are random real-symmetric
$N\times N$ matrices from the independent Gaussian Orthogonal
Ensembles described by the correlation functions $\langle
(H_{0})_{nm}\rangle=\langle V_{nm} \rangle=0$:
\begin{eqnarray}
\label{RMT-corr} & &\langle
(H_{0})_{nm}(H_{0})_{n'm'}\rangle=N(\delta/\pi)^{2}\,[\delta_{nn'}\delta_{mm'}+
\delta_{nm'}\delta_{mn'}],\\ \nonumber & & \langle V_{nm}V_{n'm
'}\rangle=(\Gamma\delta/\pi)\,[\delta_{nn'}\delta_{mm'}+
\delta_{nm'}\delta_{mn'}],
\end{eqnarray}
with $\delta=1/(\nu\, Vol)$ being the mean level spacing.

In this RMT limit {\it all} the disordered and chaotic systems
described by a real, spin-rorational invariant Hamiltonian are
believed to have a universal behavior in the limit
$L,N\rightarrow\infty$ which is characterized by only two parameters
$\delta$ and $\Gamma$ and one dimensionless function $\varphi(t)$.

Let us first consider the disorder-averaged  $G^{K}$ in
the non-crossing approximation. Since there are no vector vertices
in the present model, the ladder diagrams (Lose Diffuson) similar to 
Fig.2 make
the main contribution. The corresponding diagrams are shown in
Fig.16 where the wavy line describes the $\langle V V \rangle$
correlator, Eq.(\ref{Gamma}).
 \begin{figure}[h]
%\fbox{\vtop to8cm{\vss\hsize=.8\hsize\vglue
%0cm\hspace{0.01\hsize}\hskip -.5cm \hspace{0.1\hsize}\vss}}
\includegraphics[width=90mm]{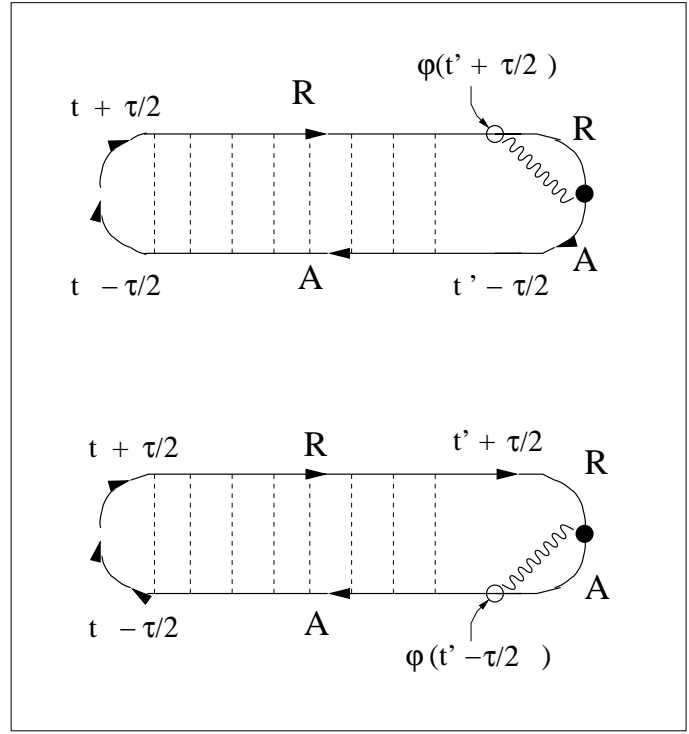}
\caption{The Loose Diffuson diagrams for the absorption rate: the dotted 
lines represent the $\langle U U\rangle$ correlation function while the 
wavy line corresponds to the $\langle V V \rangle$ correlator.}
\label{zero-loop}
\end{figure}
The condition
\begin{equation}
\label{Gtau} \Gamma\tau\ll 1
\end{equation}
allows to keep only the linear in $\hat{V}$ term in the expansion
Eq.(\ref{approxi}) for the $\langle G^{R/A}\rangle $ function
attached to the RA-junction. Note that the diagrams Fig.16a,b have
opposite signs because $\int d\xi\,(\xi\pm i/2\tau)^{-1}=\mp \pi
i$. The result of calculation of both diagrams is:
\begin{eqnarray}
\label{F16}  \delta G^{K}&=&(2\pi\nu\tau)^{2}
(i\pi\nu)(\Gamma/\pi\nu)\,D_{\eta}(t,t')\\ \nonumber &\times&
[\varphi(t'+\eta/2)-\varphi(t'-\eta/2)]^{2}\,h_{0}(\eta).
\end{eqnarray}
so that
\begin{equation}
\label{h-sol} h_{t}(\eta)=\left(1-\tilde{D}_{\eta}(t,t')\,\Gamma\,
[\varphi(t'+\eta/2)-\varphi(t'-\eta/2)]^{2}\right)\,h_{0}(\eta),
\end{equation}
where the integration over $t'$ is assumed; $h_{0}(\eta)$ is
determined by Eq.(\ref{h}), and $\tilde{D}_{\eta}(t,t')=2\pi\nu
\tau^{2}\,D_{\eta}(t,t';{\bf k}=0)$ is given by
Eq.(\ref{time-Diff}) with $D_{0}\rightarrow\Gamma$. One can check
using this equation that $h_{t}(\eta)$ obeys the equation:
\begin{equation}\label{Eq-h}
\left\{\partial_{t}+\Gamma\,\left[\varphi(t+\eta/2)-\varphi(t-\eta/2)
\right]^{2} \right\}\,h_{t}(\eta)=0.
\end{equation}
This allows to give an explicit solution for $h_{t}(\eta)$:
\begin{equation}
\label{explic}
h_{t}(\eta)=h_{0}(\eta)\,\exp\left\{-\Gamma\int_{0}^{t}\left[
\varphi(\zeta+\eta/2)-\varphi(\zeta-\eta/2)\right]^{2}\,d\zeta\right\}
\end{equation}
Note that a nontrivial dynamics of $h_{t}(\eta)$ and thus that of
$f(\varepsilon;t)$ is hidden in the exponential factor which is a
purely classical object. The Fermi statistics of electrons
considered here is taken into account by an initial distribution
$h_{0}(\eta)$. At $\eta\omega\ll 1$ where $\omega$ is the typical
frequency of oscillations in $\varphi(t)$, or equivalently at the
energy resolution $\delta\varepsilon\gg \omega$, one can
approximate $\varphi(\zeta+\eta/2)-\varphi(\zeta-\eta/2)\approx
(\partial_{t}\varphi)\, \eta$. Then the exponent in
Eq.(\ref{explic}) reduces to
$\exp\left[-t\,\Gamma\,\overline{(\partial_{t}\varphi)^{2}}\,\eta^{2}\right]$
which corresponds to Eq.(\ref{Eq-h}) of the form:
\begin{equation}\label{Eq-h-diff}
\left\{\partial_{t}+D_{E}\,\eta^{2} \right\}\,h_{t}(\eta)=0,\;\;\;
\left\{\partial_{t}-D_{E}\,\partial^{2}_{\varepsilon}
\right\}\,f(\varepsilon,t)=0.
\end{equation}
In this approximation we come to the diffusion equation for the
energy distribution function with the energy diffusion
coefficient:
\begin{equation}
\label{D} D_{E}=\Gamma\overline{(\partial_{t}\varphi)^{2}}\sim
\Gamma\,\omega^{2}.
\end{equation}
For a harmonic perturbation $\varphi(t)=\cos(\omega t)$ the form
of energy diffusion coefficient allows a simple interpretation.
This is a random walk in the energy space 
due to a sequential absorption or emission of photons with the energy
$\hbar\omega$.
Then the size of an
elementary step is $\pm \hbar\omega$ and the rate of making steps
is $\Gamma/\hbar$. Now it is clear that the condition
Eq.(\ref{Gtau}) that allows to use the approximation
Eq.(\ref{approxi}) has a physical meaning of a condition to
absorb/emit {\it at most} one photon during the time of elastic
mean free path. The opposite condition would mean that many
photons can be absorbed/emitted during the time $\tau$ which
implies inelastic processes being stronger than the elastic ones.

Using the diffusion equation we express
$\partial_{t}h_{t}(\eta)=-D_{E}\,\eta^{2}\,h_{t}(\eta)$. Then
Eq.(\ref{W}) gives for the net absorption rate:
\begin{equation}
\label{netabs} W(t)=-i\pi\,(D_{E}/\delta)\,\lim_{\eta\rightarrow
0}\partial_{\eta}(\eta^{2}h_{t}(\eta)).
\end{equation}
It is important that
\begin{equation}
\label{lim} h_{t}\approx \frac{i}{\pi\eta},\;\;\;\;\eta\rightarrow
0.
\end{equation}
for {\it all} $h_{t}(\eta)$ corresponding to the Fermi-like
energy distribution function $f(\varepsilon;t)<1$ with
$f(\varepsilon\rightarrow+\infty;t)=0$,
$f(\varepsilon\rightarrow-\infty;t)=1$. Thus the existence  of the
Fermi sea makes the net absorption rate non-zero despite the
diffusion character of the energy distribution dynamics:
\begin{equation}
\label{W-D} W(t)=W_{0}=\frac{D_{E}}{\delta}.
\end{equation}

\section{Quantum correction to absorption rate.}
We see that the non-crossing, or zero-loop, approximation leads to
the classical picture of a time-independent absorption rate, the
so called {\it Ohmic absorption}. This is in line with the fact
established below that the same approximation leads to the
classical Drude conductivity. Let us go beyond and consider the
diagrams for $G^{K}$ that contain one crossing or one
Cooperon loop. Different ways of presenting such a diagram are
shown in Fig.17.
\begin{figure}[h]
%\fbox{\vtop to10cm{\vss\hsize=.9\hsize\vglue
%0cm\hspace{0.01\hsize}\hskip -.5cm \hspace{0.1\hsize}\vss}}
\includegraphics[width=90mm]{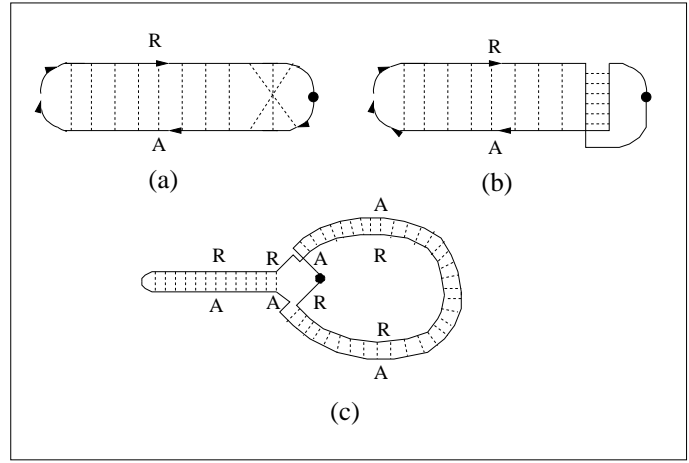}
\caption{ Diagrams for weak dynamic localization.} \label{WDLoc}
\end{figure}
This is essentially the same fan diagram with the Loose Diffuson
attached to it. However, because the perturbation does not contain
a vector vertex, one should take a special care about the Hikami
box. It is shown in Fig.18.
 \begin{figure}[h]
%\fbox{\vtop to9cm{\vss\hsize=.9\hsize\vglue
%0cm\hspace{0.01\hsize}\hskip -.5cm \hspace{0.1\hsize}\vss}}
\includegraphics[width=80mm]{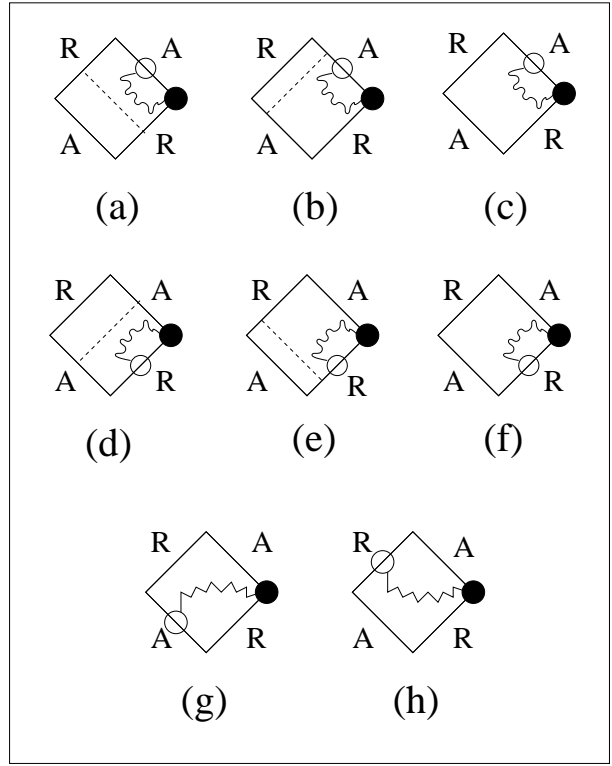}
\caption{Hikami box for a scalar vertex.} \label{Hik-box}
\end{figure}
Here a new element -- besides the correlation between two
$\hat{V}$ vertices that first appeared in Fig.16-- is the dotted
line installed between two retarded or two advanced Green's
functions. The rules of such an installation are (i) non-crossing
of dotted-dotted and dotted-wavy lines and (ii) the presence of
both retarded and advanced functions in each "cell" separated by
dotted or wavy lines (otherwise the $\xi$-integral zero). One can
see that the sum of three diagrams in Fig.18a-c and in Fig.18d-f
are zero. The remaining diagrams Fig.18g and Fig.18h have opposite
signs so that the R-A junction and the $\hat{V}(t)$ vertices
(denoted by an open circle in Fig.18) that appear as the first
term of expansion of $\langle G^{R/A}\rangle$ compose a
combination
\begin{eqnarray}
\label{comb} &\Gamma&(\varphi(t'+\eta/2)-\varphi(t'-\eta/2))\,
(\varphi(t'-t_{1})-\varphi(t'-t_{1}-\eta))\,h_{0}(\eta)\\
\nonumber &\approx&
\Gamma\,\partial_{t'}\varphi(t')\,\partial_{t'}\varphi(t'-t_{1})\,\eta^{2}
h_{0}(\eta).
\end{eqnarray}
The contribution to the absorption rate  that comes from the
diagram of Fig.17 is computed with the help of Eq.(\ref{W}). The
limit $\lim_{\eta\rightarrow 0}\partial_{\eta}\eta^{2}h_{0}(\eta)$
results in a universal constant $i/\pi$, and the time derivative
$\partial_{t}$ of the Loose Diffuson gives a $\delta(t-t')$
function in the leading order in $\hat{V}$. So we obtain a quantum
correction to the absorption rate \cite{BSK}:
\begin{equation}
\label{WDL} \frac{\delta W(t)}{W_{0}}=\frac{\Gamma\delta}{\pi
D_{E} }\,\int_{0}^{t}\partial_{t}\varphi(t)\partial_{t}
\varphi(t-t_{1})\,\tilde{C}_{t-t_{1}/2}(t_{1},-t_{1})\,dt_{1}.
\end{equation}
where the limits of integration are fixed by an assumption that a
time-dependent perturbation has been switched on at $t=0$;
$C_{t}(\eta,\eta';{\bf k})$ is the Cooperon, and
\begin{equation}
\label{tilde-Ck}
\tilde{C}_{t}(\eta,\eta')=(2\pi\nu\tau^{2})\,\frac{1}{Vol}\sum_{{\bf
k }}C_{t}(\eta,\eta';{\bf k}).
\end{equation}
Here we have to make some remarks. Eq.(\ref{WDL}) is general and
gives the first quantum correction to the absorption rate in a
disordered or chaotic system of any geometry. However, the
quantity $\tilde{C}_{t}(\eta,\eta')$ depends on the type of
perturbation and on the system geometry. For a closed system of
finite volume $Vol$ there always exists a regime where geometry of
the system does not play any role. This is the limit where the
typical frequency $\omega$ of perturbation is much smaller than
the so called Thouless energy $E_{Th}$. In disordered systems with
the diffusion character of electron transport, the Thouless energy
coincides with the gap between the {\it zero diffusion mode} that
corresponds to ${\bf k}=0$ and the first mode of dimensional
quantization that corresponds to $k\sim 2\pi/L $ , where $L$ is
the {\it largest} of the system sizes. In this case it is of the
order of the inverse diffusion time $E_{Th}\sim 1/\tau_{D}$. For
such small frequencies one can neglect the higher modes with
nonzero values of ${\bf k}$ and consider the zero diffusion mode
with ${\bf k}=0$ that always exists in closed systems. In this
{\it zero-mode}, or {\it ergodic} limit the actual shape of the
system does not matter at all. One can show that this is exactly
the limit where the results obtained using the Hamiltonian
Eq.(\ref{Ham}) are equivalent to the results obtained starting
from the random matrix theory Eq.(\ref{RMT}).

However, even in the ergodic limit the results for the Cooperon
depend on the {\it topology} of the system. The equation for the
Cooperon Eq.(\ref{time-Coop}) that corresponds to the {\it
non-single connected} toplogy of a ring and a {\it global
vector-potential} perturbation differs from that for the scalar
potential perturbations with local correlations in space described
by  Eqs.(\ref{Ham}),(\ref{Gamma}).

To see the difference we re-derive the Cooperon for the case of a
scalar potential perturbation. The corresponding diagrams
analogous to those shown in Fog.11 for the case of the global
vector-potential perturbation are given in Fig.19.
\begin{figure}[h]
%\fbox{\vtop to7cm{\vss\hsize=.9\hsize\vglue
%0cm\hspace{0.01\hsize}\hskip -.5cm \hspace{0.1\hsize}\vss}}
\includegraphics[width=110mm]{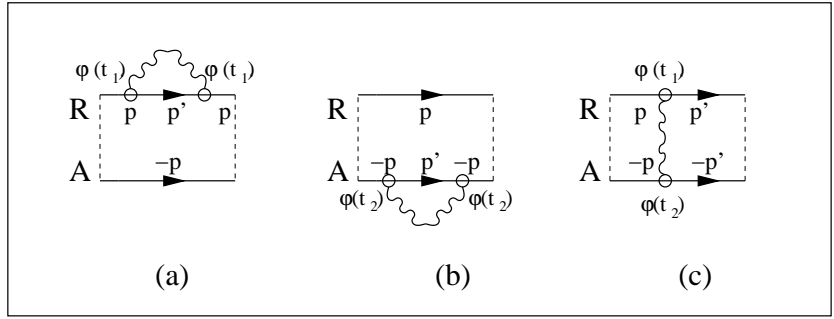}
\caption{The ac scalar potential corrections to the inverse
Cooperon.} \label{V-Coop}
\end{figure}
As a result of calculation of the corresponding $\xi$-integrals we
obtain an equation (in the ergodic limit $\bf k$=0):
\begin{equation}
\label{V-time-Coop}
\left\{2\frac{\partial}{\partial\eta}+\Gamma\left[\varphi
(t+\eta/2) -\varphi(t-\eta/2)\right]^{2}
\right\}\,\tilde{C}_{t}(\eta,\eta')=2 \delta(\eta-\eta').
\end{equation}
This equation contains the {\it difference} $\varphi (t+\eta/2)
-\varphi(t-\eta/2)$ instead of the sum in Eq.(\ref{time-Coop}).
One can show \cite{Kanzieper} that the time-dependent random matrix 
theory that
reproduces the Cooperon with the sum $\varphi (t+\eta/2)
+\varphi(t-\eta/2)$ corresponds to the random perturbation matrix
$\hat{V}$ which is pure imaginary {\it anti}-symmetric, rather
than the real symmetric as in Eq.(\ref{RMT}).

The corresponding equation for a time-dependent Diffuson appears
to be the same as Eq.(\ref{time-Diff}):
\begin{equation}
\label{V-time-Diff} \left\{\frac{\partial}{\partial
t}+\Gamma\left[\varphi (t+\eta/2) -\varphi(t-\eta/2)\right]^{2}
\right\}\,\tilde{D}_{\eta}(t,t')=\delta(t-t').
\end{equation}
To complete the solution Eq.(\ref{WDL}) we give a solution to
Eq.(\ref{V-time-Coop}) which is valid for an {\it arbitrary} time
dependence of $\varphi(t)$:
\begin{eqnarray}
\label{V-sol} \tilde{C}_{t}(\eta,\eta')=\theta(\eta-\eta')\,
\exp\left\{-\frac{\Gamma}{2}\int_{\eta'}^{\eta}\left[
\varphi(t+\zeta/2)-\varphi(t-\zeta/2)\right]^{2}\,d\zeta \right\}.
\end{eqnarray}

\section{Weak dynamic localization and no-dephasing points.}
In this section we concentrate on the {\it oscillating} time
dependence of perturbation with $\overline{\varphi(t)}=0$ and
$\overline{\varphi^{2}(t)}=1$. The simplest example is that of a
{\it white-noise} with $\overline{\varphi(t)\varphi(t')}=0$ for
$t\neq t'$. Then Eq.(\ref{V-sol}) takes the form:
\begin{equation}
\label{C-noise} \tilde{C}_{t}(\eta,\eta')=\theta(\eta-\eta')\,
\exp\left[-\Gamma\,(\eta-\eta')\right].
\end{equation}
The negative exponential factor describes {\it dephasing} by a
white noise perturbation. Because of the factor 
$\partial_{t}\varphi(t)\partial_{t}\varphi(t-t_{1})$ 
the effective range of integration in Eq.(\ref{WDL}) is  of
the order of the correlation time $\tau_{0}\ll 1/\Gamma$. This makes the
correction to the absorption rate vanishing $\delta
W(t)/W_{0}\rightarrow 0$ in the white-noise limit
$\tau_{0}\rightarrow 0$.

Now let us consider the simplest example of an ${\it
infinite-range}$ time correlations. This is the case of a harmonic
perturbation $\varphi(t)=\cos(\omega t)$. Then for
$\omega|\eta-\eta'|\gg 1$ the Cooperon takes again the form
Eq.(\ref{C-noise}) but with the {\it time-dependent} dephasing
rate:
\begin{equation}
\label{t-deph} \Gamma_{t}=\Gamma\,\sin^{2}\omega t
\end{equation}
One can see that there are certain moments of time $\omega
t_{n}=\pi n$ ($n={\rm integer}$) where the dephasing rate is zero.
We will refer to these points $t_{n}$ as {\it no-dephasing}
points \cite{Wang}.

This remarkable phenomenon is generic to harmonic perturbation and
is also present for the case of an ac vector-potential considered
in the previous section. Its physical meaning is quite simple. Consider 
again a pair
of electron trajectories with a closed loop shown in Fig.8. In the
presence of a time-dependent vector-potential ${\bf A}(t)$ the
phase difference between these two trajectories is no longer zero
but a random quantity:
\begin{equation}
\label{phase} \delta\Phi=\int_{0}^{{\cal T}}{\bf A}(t')\,[{\bf
v}_{1}(t')-{\bf v}_{2}(t')]\,dt',
\end{equation}
where ${\bf v}_{1,2}(t)$ are electron velocities at a time $t$ and
${\cal T}$ is the time it takes for an electron to make a loop
(the traversing time). For loops with the opposite directions of
traversing ${\bf v}_{1}(t')=-{\bf v}_{2}({\cal T }-t')$ [see
Fig.20]. 
\begin{figure}[h]
%\fbox{\vtop to9cm{\vss\hsize=.9\hsize\vglue
%0cm\hspace{0.01\hsize}\hskip -.5cm \hspace{0.1\hsize}\vss}}
\includegraphics[width=90mm]{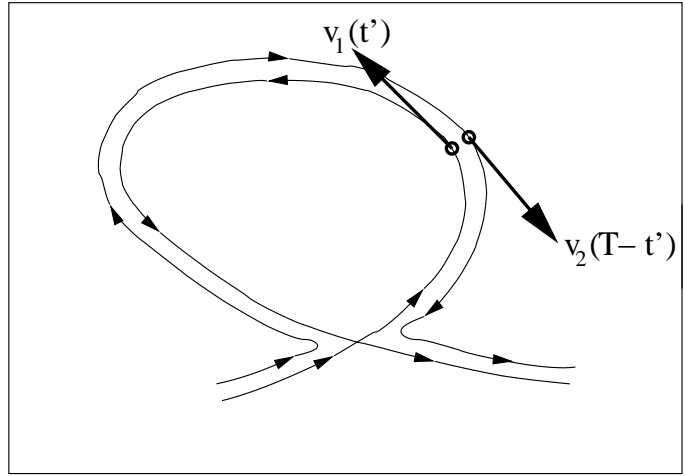}
\caption{Trajectories with loops traversed in opposite directions and 
synchronization.} \label{sel-traj}
\end{figure}
Then we obtain
\begin{equation}
\label{phase-2} \delta\Phi=\int_{-{\cal T}/2}^{{\cal T}/2}[{\bf
A}({\cal T}/2+t')+{\bf A}({\cal T}/2-t')]\,{\bf v}_{1}(t'+{\cal
T}/2)\,dt'
\end{equation}
Now assuming the period of oscillations of ${\bf A}(t)$ to be
large compared to the velocity correlation time $\tau$ we obtain
after averaging over ${\bf v}_{1}$:
\begin{equation}
\label{phase-3} \langle (\delta \Phi)^{2}\rangle=D_{0}\int_{-{\cal
T}/2}^{{\cal T}/2}[{\bf A}({\cal T}/2+t')+{\bf A}({\cal
T}/2-t')]^{2}\,dt'.
\end{equation}
We see that in the presence of a harmonic vector-potential
${\bf A}(t)\propto \sin\omega t$ the phase difference $\delta
\Phi=0$ vanishes for {\it all} the loop trajectories with the
traversing time equal to the integer period of the time-dependent
perturbation:
\begin{equation}
\label{trav-time} {\cal T}=\frac{2\pi n}{\omega},\;\;\;\;n=1,2,3...
\end{equation}
Such loop trajectories with the traversing time {\it synchronized}
with the period of perturbation play a crucial role in all the
quantum coherence (interference) phenomena in disordered and
chaotic systems (their effect on the universal conductance fluctuations
has been studied in Ref.\cite{Wang}). Any periodic in time perturbation
leads to a {\it selection of trajectories}: for 
strong enough  
perturbation all 
trajectories that do not obey Eq.(\ref{trav-time}) do not interfere 
because of
the random phase difference $\delta\Phi\gg 1$.

In the particular case of the quantum corrections to the energy
absorption rate Eq.(\ref{WDL}) the time-dependent Cooperon
\begin{equation}
\label{G-Coop}
\tilde{C}_{t-t_{1}/2}(t_{1},-t_{1})=\exp\left[-2t_{1}\Gamma_{t-t_{1}/2}\right]
\end{equation}
is small everywhere except for the vicinity of no-dephasing points
\begin{equation}
\label{ndp} \omega t_{1}^{(n)}=2\omega t -2\pi n,\;\;\;n=0,\pm
1,\pm 2...
\end{equation}
For $\Gamma t_{1}\sim \Gamma t\gg 1$ one can expand
$\Gamma_{t-t_{1}/2}\approx (\omega/2)^{2}\, \Gamma
\,(t_{1}-t_{1}^{(n)})^{2}$ in the vicinity of no-dephasing points,
do the Gaussian integration over $\zeta= (t_{1}-t_{1}^{(n)})$ and
put $t_{1}=t_{1}^{(n)}$ elsewhere. In particular
$\partial_{t}\varphi(t-t_{1}^{(n)})=-\partial_{t}\varphi(t)$. Then
we arrive at:
\begin{equation}
\label{WLC} \frac{\delta W(t)}{W_{0}}= -
\frac{\delta}{\pi}\,(2\sin^{2}\omega
t)\,\sum_{n}\int_{-\infty}^{+\infty}d\zeta\,e^{-D_{E}\zeta^{2}t_{1}^{(n)}}.
\end{equation}
Replacing $\sum_{n}$ by the integral $(\omega/2\pi)\int_{0}^{t}
dt_{1}$ and averaging over time intervals much larger than
$2\pi/\omega$ we finally get:
\begin{equation}
\label{sqrt}
 \frac{W(t)}{W_{0}}=1-\frac{\delta\omega}{\pi}\int_{0}^{t}dt_{1}\int_{-\infty}^{+\infty}
 \frac{d\zeta}{2\pi}\,e^{-D_{E}\zeta^{2}t_{1}}=
 1-\sqrt{\frac{t}{t_{*}}},
\end{equation}
where
\begin{equation}
\label{t*} t_{*}=\frac{\pi^{3}\Gamma}{2\delta^{2}}.
\end{equation}
We have obtained a remarkable result that the absolute value of
the  quantum correction to the absorption rate is growing with
time. This is the consequence of the existence of no-dephasing
points, as otherwise the exponentially decaying Cooperon would
lead to a saturation of the integral over $t_{1}$ in
Eq.(\ref{WDL}) at large times $t$. The negative sign of the
correction implies that the absorption rate, or energy diffusion
coefficient decreases with time. This phenomenon can be called
"weak dynamic localization" in full analogy with the weak Anderson
localization when the diffusion coefficient in space is decreasing
with the system size. 
It has been first discovered \cite{Casati79,Casati90,Altland} for a 
simple quantum system -- 
quantum rotor
subject to the periodic $\delta$-function perturbation. 
At $t\sim t_{*}$ the quantum correction is
of the order of the classical Ohmic absorption, and we can expect
the strong {\it dynamic localization } to occur.

Eq.(\ref{sqrt}) suggests a more precise relationship between the
dynamic localization for a quantum system  in the ergodic
({\it zero-dimensional}) limit subject to a harmonic perturbation and
the Anderson localization in a {\it quasi-one dimensional} 
disordered wire.
We observe that
\begin{equation}
\label{obser}
\partial_{t}W(t)\propto
\int_{-\infty}^{+\infty}\frac{d\zeta}{2\pi}\,e^{D_{E}\zeta^{2}\,t}=
\int_{-\infty}^{\infty}\frac{d\omega'}{2\pi}\,e^{-i\omega' t }
\int_{-\infty}^{+\infty} \frac{d\zeta}{2\pi}\,\frac{1}
{D_{E}\zeta^{2}-i\omega'}.
\end{equation}
At the same time, according to Eq.(\ref{WL})the correction the
(complex) frequency dependent conductivity of an infinite
quasi-one dimensional wire is proportional to:
\begin{equation}
\label{fdcon}
\delta\sigma(\omega)=\int_{-\infty}^{+\infty}\frac{dk}{2\pi}\,
\frac{1}{D_{0}k^{2}-i\omega}.
\end{equation}
We see that the deviation from the no-dephasing points
$\zeta=t_{1}-t_{1}^{(n)}$ is an analogue of the momentum $k$. Then
with a suitable choice of a time and frequency scale one obtains:
\begin{equation}
\label{AL-DL}
\frac{W(t)}{W_{0}}=\int_{-\infty}^{+\infty}\frac{d\omega'}{2\pi}\,
\frac{e^{-i\omega'
t}}{(-i\omega'+0)}\,\left(\frac{\sigma(\omega')}{\sigma_{0}}\right).
\end{equation}
Note that Eq.(\ref{AL-DL}) is non-trivial as it establishes a 
relationship between an essentially non-equilibrium property of a 
zero-dimensional system and an equilibrium property of a quasi-one 
dimensional system. 

This relationship can be proven for any diagram with an 
arbitrary
number of the Diffuson-Cooperon loops (for the case of two loops see 
Ref.\cite{SBK}) and we conjecture that it is
{\it exact} in the ergodic (or random matrix) regime for $t\Gamma,
t\omega\gg 1 $.

Eq.(\ref{AL-DL}) helps to establish a character of decay of the
absorption rate $W(t)$ for $t\gg t_{*}$. To this end we recall the
Mott-Bereznskii formula for the frequency-dependent conductivity
in the localized regime \cite{Mott}:
\begin{equation}
\label{Ber} \sigma(\omega)\propto \omega^{2}\,\ln^{2}\omega.
\end{equation}
Then using Eq.(\ref{AL-DL}) one obtains the corresponding 
absorption rate 
$W(t)$ at $t\gg
t_{*}$:
\begin{equation}
\label{tail} W(t)\propto \frac{\ln t}{t^{2}}.
\end{equation}
 \begin{figure}[h]
%\fbox{\vtop to9cm{\vss\hsize=.9\hsize\vglue
%0cm\hspace{0.01\hsize}\hskip -.5cm \hspace{0.1\hsize}\vss}}
\includegraphics[width=100mm]{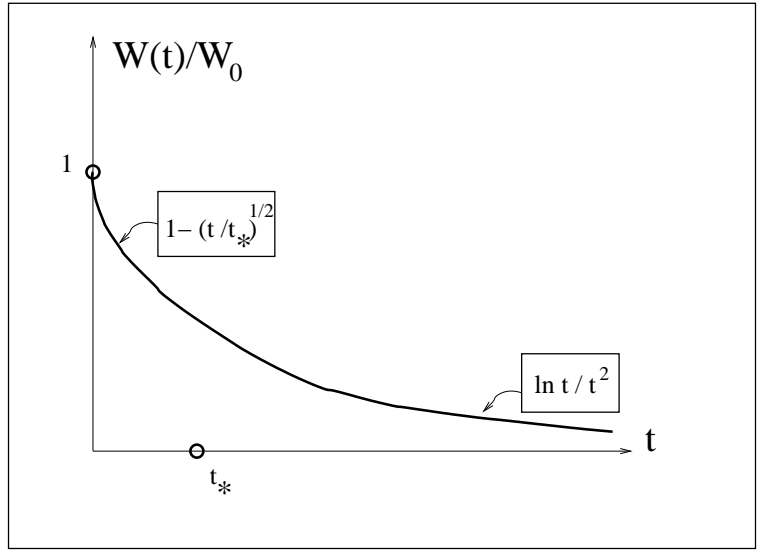}
\caption{The time dependence of the absorption rate.}
\label{Dynamic loclization}
\end{figure}
\section{Conclusion and open questions}
The goal of this course was to give a unified picture and a
unified theoretical tool to consider different quantum coherence
effects in disordered metals. The unified picture is that of
interference of electron trajectories with loops traversed in the
opposite directions. The corresponding theoretical machinery is
the Diffuson-Cooperon diagrammatic technique. We have demonstrated
that even the first quantum correction diagram with the Cooperon
loop can describe such  different and nontrivial phenomena as weak
Anderson localization, quantum rectification and dynamic
localization in the energy space. We did not try to give a review
of the development in the field but rather to concentrate on few
important examples and to demonstrate how does the machinery work
in different cases. That is why many related issues have not been
discussed and the corresponding works have not been cited
properly. We apologize for that.

There are few open problems that are related with the main
subjects of this school and in our opinion are warrant a study.
This is first of all a unified theory of energy absorption where
both the Zener transitions picture \cite{Thouless}  and the sequential 
photon
absorption picture are incorporated. The suitable theoretical tool
for that is believed to be a nonlinear sigma-model in the Keldysh
representation derived in Refs.\cite{BSK}. The perturbative
treatment of this field theory reproduces the diagrammatic
technique discussed in this course, and the non-perturbative
consideration in the region $\Gamma,\omega<\delta$ should give the
results obtained in the framework of the Zener transitions
picture.

Another possible direction is the role of interaction for the
quantum rectification and dynamic localization. Some of the
interaction effects in dynamic localization have been recently
considered in Refs.\cite{Bas,Bas-Kr,Kra} using the Fermi Golden
Rule approximation. However, an interesting regime of localization
in the Fock space \cite{LevKam} where the Fermi Golden Rule does not 
apply 
is
awaiting an investigation.

\end{document}